\newcommand{\rcor}{$R_{\rm cr}$}
\newcommand{\rbar}{$R_{\rm bar}$}
\newcommand{\omegabar}{$\Omega_{\rm bar}$}
\newcommand{\sbar}{$S_{\rm bar}$}
\newcommand{\rr}{${\cal R}$}
\newcommand{\vcirc}{$V_{\rm circ}$}
\newcommand{\msun}{M_{\odot}}
\newcommand{\rpetro}{$R_{\rm Petro}$}

\documentclass{aa}
\usepackage{newtxtext,newtxmath}
\usepackage{float}
\usepackage{rotating}
\usepackage{graphicx}
\usepackage{longtable}  
\usepackage{pdflscape}
\usepackage{multicol}

\usepackage[T1]{fontenc}
\usepackage{ae,aecompl}

\usepackage{graphicx}	
\usepackage{amsmath}	
\usepackage{amssymb}	

\usepackage[dvipsnames,usenames,table]{xcolor}
\usepackage{longtable}

\begin{document}
\title{Relations among structural parameters in barred galaxies with a direct measurement of bar pattern speed}

   \author{Virginia Cuomo
          \inst{1,2}
          \and J. Alfonso L. Aguerri\inst{3,4}
          \and Enrico Maria Corsini \inst{1,5}
          \and Victor P. Debattista\inst{6}
          }

   \institute{Dipartimento di Fisica e Astronomia ”G. Galilei”, Università di Padova, vicolo dell'Osservatorio 3, I-35122 Padova, Italy\\
              \email{virginia.cuomo@phd.unipd.it}
         \and
             Instituto de Astronom\'ia y Ciencias Planetarias, Universidad de Atacama, Avenida Copayapu 485, Copiap\'o, Chile
         \and
             Departamento de Astrof\'isica, Universidad de La Laguna, Avenida Astrof\'isico Francisco S\'anchez s/n, E-38206 La Laguna, Tenerife, Spain
        \and
            Instituto de Astrof\'isica de Canarias, calle V\'ia L\'actea s/n, E-38205 La Laguna, Tenerife, Spain
        \and
            INAF - Osservatorio Astronomico di Padova, vicolo dell'Osservatorio 2, I-35122 Padova, Italy
        \and
           Jeremiah Horrocks Institute, University of Central Lancashire, PR1 2HE Preston, UK
         }
    
\date{Accepted XXX. Received YYY; in original form ZZZ}



\abstract{We investigate the relations between the properties of bars and their host galaxies in a sample of 77 nearby barred galaxies, spanning a wide range of morphological types and luminosities, with 34 SB0-SBa and 43 SBab-SBc galaxies. The sample includes all the galaxies with reliable direct measurement of their bar pattern speed based on long-slit or integral-field stellar spectroscopy using the Tremaine-Weinberg method. We limited our analysis to the galaxies with a relatively small relative error on the bar pattern speed ($\leq50$ per cent) and not hosting an ultrafast bar. For each galaxy, we collected the radius, strength, pattern speed, corotation radius, and rotation rate for the bar and we also collected the Hubble type and absolute SDSS $r$-band magnitude. We also used literature bulge-to-total luminosity ratio for a subsample of 53 galaxies with an available photometric decomposition. We confirmed earlier observational findings that longer bars rotate with lower bar pattern speeds, shorter bars are weaker, and bars with a small bar rotation rate rotate with higher bar pattern speeds and have smaller corotation radii. In addition, we found that stronger bars rotate with lower bar pattern speeds, as predicted from the interchange of angular momentum during bar evolution, which in turn may depend on different galaxy properties. Moreover, we report that brighter galaxies host longer bars, which rotate with lower bar pattern speeds and have larger corotation radii. This result is in agreement with a scenario of downsizing in bar formation, if more massive galaxies formed earlier and had sufficient time to slow down, grow in length, and push corotation outwards.}

\keywords{galaxies: kinematics and dynamics --- galaxies: formation --- galaxies: evolution --- galaxies: fundamental parameters --- galaxies: structure}

\titlerunning{Relations among parameters in barred galaxies with a direct measurement of bar pattern speed}
\maketitle



\section{Introduction}
\label{sec:bar_properties_introduction}

Bars are common in the local Universe across a wide range of galaxy morphologies \citep[ e.g.,][]{Aguerri2009,Buta2015}, luminosities \citep[ e.g.,][]{MendezAbreu2010,erwin2018}, and environments \citep[e.g.,][]{mendezabreu2012,lin2014}. The photometric, kinematic, and dynamical properties of bars have been studied extensively \citep[e.g.,][]{Debattista2005,Gadotti2011}: their formation mechanisms and evolutionary processes include the interchange of angular momentum \citep[e.g.,][]{Debattista2000,athanassoula2003,Berentzen2004,Sellwood2006,Sellwood2006b,villavargas2010,Athanassoula2013,Lokas2014}.

According to their overall shape and in addition to the classical bar morphology characterised by a smooth light distribution, bar-like features can have also an ansae-type morphology with a light concentration at each end \citep[e.g.,][]{Laurikainen2007, MartinezValpuesta2007}. Bars can be roughly divided into `flat' and `exponential' based on their surface brightness radial profile. A flat bar has a flatter profile than the surrounding disc, whereas the profile of an exponential bar is more similar to that of the disc \citep{Elmegreen1985, Elmegreen1996a}. Flat bars are more typical of early-type (ETBGs, with Hubble stage ranging between ${\rm T}=1$ to 5) rather than late-type barred galaxies (LTBGs, with ${\rm T}$ between 5 and 7) and can exhibit isophotal twists \citep{Elmegreen1985}. In particular, exponential profiles are typical in LTBGs with ${\rm T}\geq5$, while flat ones in systems with $T<5$, as shown by \cite{DiazGarcia2016} using near-infrared imaging of the S$^4$G survey \citep{Sheth2010}. Moreover, the profiles change according to the mass, as studied by \cite{Kruk2018} using the GalaxyZoo project \citep{Lintott2008}. In fact, low-mass, disc-dominated galaxies have bars with an almost exponential light profile, while high-mass galaxies with a prominent bulge have bars with a flat profile.

The primary parameters describing a bar are its radius, $R_{\rm bar}$, strength, $S_{\rm bar}$, pattern speed, $\Omega_{\rm bar}$, and rotation rate \rr. The radius and strength of the bar are structural parameters and can be recovered from the analysis of optical and/or near-infrared images, while the pattern speed and rotation rate of the bar are dynamical parameters and their determination requires kinematics. 

The relations between the properties of bars and their host galaxies have also been widely explored. Early findings include the results by \cite{Elmegreen1985}, who found that bars in ETBGs are longer than those in LTBGs, and by \cite{Athanassoula1980}, who observed that galaxies with smaller bulges also have shorter bars. \cite{Kim2016} investigated the relations between the properties of bars and those of the inner parts of their host discs based on near-infrared Spitzer data. They found that among massive galaxies ($M_\ast>10^{10}\msun$), longer bars reside in more flattened inner discs (i.e., discs with larger inner scalelengths and lower central surface brightnesses) than shorter bars. Moreover, such bars are often associated with a light deficit in the disc surrounding them. This deficit turns out to be more pronounced with longer bars and/or in galaxies with a higher bar-to-total luminosity ratio at all the explored masses ($10^9<M_\ast<10^{11}\msun$). By performing a bi-dimensional decomposition on multiband SDSS images of 700 galaxies, \cite{SanchezJanssen2013} found that barred discs tend to have fainter central surface brightness, and larger disc scalelengths $h$ than unbarred galaxies of the same stellar mass ($M_\ast>10^{10}\msun$). Barred galaxies appear to be more extended than the unbarred counterparts \citep{Erwin2019}. \cite{Consolandi2016} found that massive spiral galaxies harbour red and dead bars, which become even redder with increasing galaxy mass. \cite{Aguerri1999} found that global star formation activity is enhanced when a galaxy hosts a strong bar. Several studies have been devoted to this subject with conflicting conclusions \citep[e.g.,][]{Wang2020}. Star formation in strongly-barred galaxies resulted to be significantly lower than in unbarred galaxies \citep{Kim2017}, while an almost complete suppression of the star formation in the bar region is often associated with its increase both at the centre of the bar and in a ring just outside it, implying that the gas is redistributed by the bar \citep{James2018,Neumann2020}.

\cite{DiazGarcia2016,DiazGarcia2016b} studied the shapes of bars. In lenticular and low-mass objects ($M_\ast<10^9\msun$), bars are oval-shaped, while in early- and intermediate-type spirals ($0<{\rm T}\leq5$) bars are narrower than in later types. However, the shape of bars among ETBGs can be rounded by the presence of a bulge and/or barlens. Concerning the strength of bars, the discussion is controversial because \sbar\ strongly depends on the method used to measure it \citep{Lee2019}. \cite{DiazGarcia2016b} suggested that more massive ETBGs ($M_\ast>10^9\msun$) host stronger bars, defined assuming the bar Fourier density amplitudes. Those bars are characterised by more discy inner isophotes as well. These results agree with a scenario whereby intermediate barred galaxies (${\rm T}=5$) move in the Hubble sequence towards earlier-types, while bars trap stars from the disc and become narrower and stronger \citep{Kormendy2013}. This scenario is supported by the findings that bars in ETBGs are longer (both in physical units and relative to the disc) and have larger \sbar, as measured by the bar Fourier density amplitudes, than later types ($3<{\rm T}\leq5$). Bars in ETBGs seem to be on average stronger than those in LTBGs, even when \sbar\ is determined by visual inspection \citep{Lee2019}. On the other hand, \cite{MenendezDelmestre2007} found a similar distribution in the mean bar ellipticity, adopted as estimate of \sbar, in ETBGs and LTBGs, so that \sbar\ appears uncorrelated with the Hubble type. \cite{Erwin2005} showed that bars in ETBGs are larger than bars in LTBGs, but this is true also for relative sizes (for example when considering \rbar\ relative to the radius of the isophote with 25 mag arcsec$^{-2}$ in $B$-band $R_{25}$ or to $h$). Moreover, ETBGs present a strong correlation between \rbar\ and disc size, but this correlation disappears in LTBGs. When splitted according to a morphological classification, strong and weak bars in ETBGs differ primarily in ellipticity, while they have very similar sizes. However, strong bars in LTBGs are on average twice the size of weak bars. These conclusions were partially revised with the advent of the S$^4$G survey \citep[e.g.,][]{HerreraEndoqui2015,DiazGarcia2016,DiazGarcia2016b,Erwin2019}. In fact, \cite{DiazGarcia2016} found that bars in LTBGs (${\rm T}>5$) are unexpectedly long, with respect to both $R_{25}$ and $h$, while their physical sizes are small. The bar ellipticity decreases from ${\rm T} = 0$ towards earlier types, as found by \cite{Laurikainen2007}, whereas the maximum ellipticity remains nearly constant along the Hubble sequence (as in \citealt{Marinova2007}).  

The structural and dynamical evolution of a barred galaxy is driven by $\Omega_{\rm bar}$. This is the angular speed of the bar and determines how far from the galaxy centre the bar affects the orbits of stars. It is parametrised by the bar rotation rate ${\cal R} \equiv R_{\rm cr}/R_{\rm bar}$, which is the ratio between the corotation radius, \rcor, and the bar radius, \rbar. Stellar dynamics sets an upper limit to $\Omega_{\rm bar}$ \citep{conto1980}: the bar can extend only as far as the corotation radius \rcor\ (where \omegabar\ equals the circular frequency) and this implies ${\cal R}\geq1$. On the other hand, there is no lower limit for $\Omega_{\rm bar}$, since the bar can be much shorter than \rcor. The bar rotation rate, ${\cal R}$, distinguishes between `fast' and `slow' bars when $1\leq {\cal R} \leq 1.4$ and ${\cal R}>1.4$, respectively. The limiting value 1.4 was set by numerical simulations \citep{athanassoula92, Debattista2000} and it does not imply any specific range for $\Omega_{\rm bar}$. The case of ${\cal R}<1$ corresponds to an unstable regime for stellar orbits. Nevertheless, some example of these `ultrafast' bars have been suggested \citep{Buta2009, Aguerri2015, Guo2019}. Theoretical studies \citep{weinberg1985} and simulations \citep[e.g.,][]{Little1991,Hernquist1992,Debattista1998,Debattista2000} find an efficient and rapid slowdown of the rotation of the bar due to the dynamical friction exerted by the halo. This corresponds to an evolution of \rr\ from the fast to the slow regime.

A sustained effort has been devoted to determining \omegabar\ and \rr\ in a large number of galaxies to understand how they are related to other properties of the bar and other galaxy components. \cite{Rautiainen2008} derived \rcor\ for 38 barred galaxies. They compared the observed morphology with that predicted by a set of dynamically-motivated numerical simulations with different \omegabar\ and dark matter halo contributions. They found some weak correlations between \rr\ and galaxy morphology, with ETBGs hosting fast bars and LTBGs showing both fast or slow bars. In this method, slow bars tend to be shorter and weaker, when \sbar\ is given by the gravitational torque of the bar, with no clear trend with either galaxy luminosity or colour. \cite{Rautiainen2008} pointed out that their findings depend strongly on the adopted modelling and leave room for the possibility that the derived pattern speed in many galaxies is that of the spiral structure rather than the bar's.  They also claimed that the error estimates of model-based pattern speeds are typically smaller than those of the TW method, despite it is controversial to compare the error estimates obtained with different methodologies (see \citealt{Cuomo2019}, for a discussion). \cite{Font2017} derived \rcor\ in 68 spirals using the phase-reversals of gas streaming motions \citep[see also][for a validation of this method]{Beckman2018}. Most of these bars are fast: \rr\ increases from ${\rm T}=1$ to ${\rm T}=3$ galaxies, then remains constant to ${\rm T}=7$ galaxies and drops for ${\rm T}=9$ galaxies. More massive galaxies ($M_\ast>3.2\times 10^{10}\msun$) host both strong and weak bars, when \sbar\ is given by the bar Fourier density amplitudes, and the longest bars rotate with lower \omegabar. Intermediate-mass galaxies ($3.2\times 10^{9}<M_\ast\leq3.2\times 10^{10}\msun$) host the shortest bars, covering both the slow and fast regimes. Less massive galaxies ($M_\ast<3.2\times 10^{9}\msun$) host only weak bars, which always rotate slowly and can be very short. \cite{Font2017} suggested that evolved barred galaxies are characterised by large stellar masses and slowly-rotating long bars. This is consistent with the two main scenarios for bar formation, namely internal processes or tidal interactions. Moreover, the authors reported that bars in the fast regime can rotate with very low values of \omegabar, suggesting these bars can increase their length more quickly than \rcor, while they are braked.

Hovewer, the pattern speed estimates obtained by \cite{Rautiainen2008} and \cite{Font2017} are not directly measured, contrary to the method proposed by \citet[][hereafter TW]{tw}. 

The early applications of the TW method to long-slit spectroscopic data of a small number of galaxies prevented inferring any firm conclusion about the relations between \omegabar\ (and \rr) and global galaxy properties. Recently, the TW method has been applied to integral-field spectroscopic data of a large number of galaxies.
\cite{Aguerri2015} investigated 15 strongly barred galaxies in the CALIFA survey \citep{sanchez2012,Walcher2014}. They combined their results with previous measurements based on the TW method \citep[see][for a review]{Corsini2011}, collecting a sample of 32 galaxies. For all these galaxies, they found that $1.0 \leq {\cal R} \leq 1.4$ independent of Hubble type. \cite{Aguerri2015} concluded that both ETBGs and LTBGs are consistent with hosting fast bars, but they were limited by their small sample.
\cite{Guo2019} performed a similar analysis on a sample of 51 galaxies from the MaNGA survey \citep{bundy2015}. They found that larger bars are stronger, when \sbar\ is measured by the bar Fourier density amplitudes, but they were unable to find any trend between \rr\ and stellar age, metallicity, bar strength or dark matter fraction within an effective radius. \cite{Guo2019} argued this is due to the various factors involved in the slowdown process and angular momentum exchange. However, it should be noticed that the large uncertainties of their measurements could have heavily affected their conclusions.
Finally, \cite{Cuomo2019bis} analysed 16 weakly barred CALIFA galaxies and found that weak bars, quantitatively defined using the bar Fourier density amplitudes, have shorter \rbar\ and \rcor\ but similar \omegabar\ and \rr\ as strong bars. The fact that weak bars are fast excludes that they formed by tidal interactions \citep{martinez2017, Lokas2018}. \citet{GarmaOehmichen2019} analysed a sample of 18 galaxies from both CALIFA and MaNGA data, 13 of which were taken from \cite{Aguerri2015} and \cite{Guo2019} to test the applicability of the TW method. The authors found that longer bars present larger corotation radii and tend to rotate at lower pattern speeds. No correlation was found between \rr\ and the total stellar mass nor the Hubble type, as in previous works, but a weak correlation between \rr\ and molecular gas mass was noted. Moreover, they observed an increase in \rr\ and a decrease of \omegabar\ with an increase of the molecular gas fraction,  which they interpreted as evidence of a more efficient slowdown of the bar in gaseous discs. Finally, \omegabar\ appeared to decrease with the stellar mass, suggesting that the most massive galaxies host bigger and slower bars. However, these relations have to be considered with caution, given the small number of analysed galaxies and large errors on the measured quantities.

In this paper we revisit the full sample of galaxies with a TW-measured \omegabar. We collect, for the first time, all the measurements based on the TW method available in literature, doubling the number of analysed galaxies with respect to similar previous works \citep{Aguerri2015,Cuomo2019bis,Guo2019}. The galaxies span a wide range of \rbar, \sbar, and \omegabar\ with direct measurements, to infer possible relations between the properties of bars and their host galaxies. The paper is structured as follows. We present the galaxy sample in Sect.~\ref{sec:sample}. We collect the bar properties of the sample galaxies in Sect.~\ref{sec:bar_parameters} and tabulate them in the Appendix~\ref{appendix:a}. We present and discuss our results in Sect.~\ref{sec:bar_properties_results} and \ref{sec:bar_properties_discussion}, respectively. We summarise our conclusions in Sect.~\ref{sec:conclusions}. We adopt as cosmological parameters $\Omega_{\rm m} = 0.286$, $\Omega_{\Lambda} = 0.714$, and $H_0=69.3$ km s$^{-1}$ Mpc$^{-1}$ \citep{Hinshaw2013}.

\section{The sample}
\label{sec:sample}

We collected a sample of 104 galaxies with a direct measurement of \omegabar\ based on stellar kinematics available in the literature. Either long-slit or integral-field spectroscopy was used to obtain the mean position and mean line-of-sight (LOS) velocity of the stars across the bar needed to apply the TW method. All the sample galaxies were analysed in a consistent way and are divided into three subsamples according to their source: 

\begin{enumerate}

\item the literature subsample (see Table~\ref{tab:sample_literature}) includes 18 galaxies taken from papers with small samples \citep{Merrifield1995, Gerssen1999, Gerssen2003, Debattista2002, Aguerri2003, Corsini2003, Corsini2007, Debattista2004, Treuthardt2007,Cuomo2019}, applying the TW method to a single or a small number of objects observed with different telescopes and instruments \citep[see][for an almost complete review]{Corsini2011}. Although the Hubble types of this subsample run from SB0 to SBbc, the majority of the galaxies ($\sim 85$ per cent) are classified as SB0 or SBa. This reflects a selection bias of the early applications of the TW method, when ETBGs were preferred because it was more straightforward to apply the TW method, since they better fullfil the constraints of the method discussed in Sect.~\ref{sec:pattern}. The redshifts are $z < 0.025$, with $\sim70$ per cent of the subsample galaxies having $z<0.01$. The absolute SDSS $r$-band magnitudes are distributed between $-18.0$ and $-22.0$ mag.

\item the CALIFA subsample (see Table~\ref{tab:sample_califa}) includes 31 galaxies taken from the CALIFA survey and has 15 strongly barred galaxies and 16 weakly barred galaxies analysed by \cite{Aguerri2015} and \cite{Cuomo2019bis}, respectively. The CALIFA survey targeted $\sim 600$ galaxies selected from the SDSS-DR7 \citep{Abazajian2009} according to their angular isophotal diameter ($D_{25}\sim 45-80$ arcsec) and redshift ($z\sim 0.005-0.03$). The galaxies of the CALIFA subsample have Hubble types ranging from SB0 to SBd, redshifts $0.005 < z < 0.03$, with most of them ($\sim85$ per cent) in the range $0.01< z < 0.02$, and absolute SDSS $r$-band magnitudes spanning between $-19.5$ and $-22.5$ mag. 

\item the MaNGA subsample (see Table~\ref{tab:sample_manga}) includes 55 galaxies taken from the MaNGA survey: 51 of them were measured by \cite{Guo2019}, while the remaining four by \cite{GarmaOehmichen2019}\footnote{ The authors repeated the analysis for the galaxies in common with \citet{Aguerri2015} and \citet{Guo2019}. For those objects we considered here the results of \citet{Aguerri2015} and \citet{Guo2019}.}. The MaNGA survey \citep{Drory2015, Yan2016, Wake2017} aims to investigate $\sim10000$ nearby galaxies from the SDSS Main Galaxy Legacy Area \citep{Abazajian2009}. The galaxies were selected to have redshifts $0.02 < z < 0.1$ and colour-based stellar masses $M_\ast> 10^9\msun$. The original 53 galaxies of \cite{Guo2019} turned out to be 51 because two  pairs of objects are repeated. The galaxies of this subsample have Hubble types ranging from SB0 to SBc, although most of them ($\sim70$ per cent) are late-type galaxies. Their redshift range reflects the mother sample distribution ($0.02 < z < 0.08$) and the galaxies of the most populated bin ($\sim55$ per cent) have $0.025 < z < 0.04$. The absolute SDSS $r$-band magnitudes are distributed between $-19.5$ mag and $-23.0$ mag. 
    
\end{enumerate}

For each galaxy in our sample, we adopted the morphological classification and redshift of its corresponding paper. If the galaxy redshift was not immediately available, we took the value given by NED\footnote{The NASA/IPAC Extragalactic Database is available at \url{https://ned.ipac.caltech.edu/}}. We calculated the absolute SDSS $r$-band magnitude $M_r$ either from the model $r$-band apparent magnitude, $m_r$, provided by the SDSS DR14, or using the apparent magnitude in a different band converted into $m_r$ using the prescriptions of \cite{Fukugita1996}. To this aim, we considered the galaxy distance from NED as obtained from the radial velocity with respect to the cosmic microwave background reference frame.

\begin{figure*}[!t]
    \centering
    \includegraphics[scale=0.7]{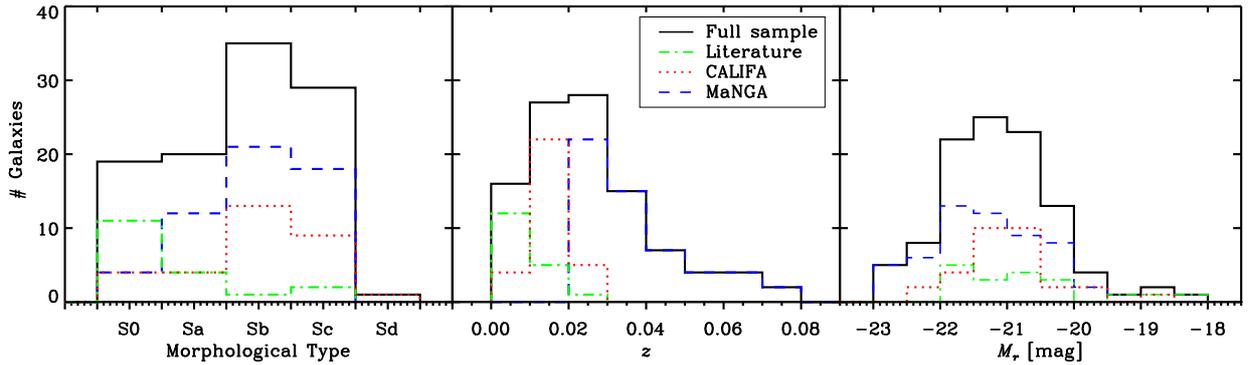}
    \caption{Distributions of Hubble types, redshifts $z$, and SDSS $r$-band absolute magnitudes $M_r$ of the full sample of 104 galaxies ({\em black solid line}), literature subsample ({\em green dot-dashed line}), CALIFA subsample ({\em red dotted line}), and MaNGA subsample ({\em blue dashed line}).}
    \label{fig:histogram_properties}
\end{figure*}

Figure~\ref{fig:histogram_properties} shows the distributions of the Hubble types, redshifts, and absolute SDSS $r$-band magnitudes of the three subsamples together with the total distribution of the entire sample. 
The literature subsample is comprised mostly of ETBGs, whereas the CALIFA and MaNGA subsamples contain a more representative number of LTBGs. The redshifts are smaller for the literature and CALIFA subsamples than for the MaNGA one. More than 90 per cent of the sample galaxies are brighter than $M_r = -20.0$ mag. Although the three subsamples show similar distributions of $M_r$, it is clear that the brightest galaxies mainly come from the MaNGA subsample and the fainter ones from the literature subsample.

Starting from the original sample of 104 galaxies, we decided to select for our analysis a subsample of 77 objects with a trustworthy measurement of \omegabar\ and \rr, as described in Sect. \ref{sec:bar_properties_uncertainties} and~\ref{sec:bar_properties_results}.

\section{Determination of the bar parameters}
\label{sec:bar_parameters}


\subsection{Bar radius}
\label{sec:abar}

The value of $R_{\rm bar}$ is a measure of the extension of the stellar orbits supporting the bar. It is quite difficult to derive because a bar does not have sharp edges and often fades smoothly into other components (like rings or spiral arms) which may affect the identification of its boundary. Several methods have been used to measure $R_{\rm bar}$. Here, we briefly review the measurement techniques adopted for the sample galaxies.

A first approach to measure $R_{\rm bar}$ is based on the direct analysis of galaxy images. A rough estimate is obtained by visual inspection of the images \citep[e.g.,][]{Merrifield1995,Treuthardt2007} or by identifying the change of slope of the surface brightness profile along the bar major axis \citep[e.g.,][]{Gerssen1999}. Two more refined techniques require the study of the radial profile of ellipticity, $\epsilon$, and position angle, PA, of the ellipses which best fit the galaxy isophotes. The galaxy isophotes show a peak of $\epsilon$ and a constant PA in the bar region, and a (generally different) constant PA in the disc region, or a non-constant behaviour (e.g., when prominent spirals arms are present). The value of $R_{\rm bar}$ is identified as the position of the maximum in the $\epsilon$ profile (e.g., \citealt{Aguerri2015,Cuomo2019bis,GarmaOehmichen2019,Guo2019}) or as the position where the PA changes by $\Delta{\rm PA} = 5\degr$ from the PA of the ellipse corresponding to the maximum in $\epsilon$ (e.g., \citealt{Aguerri2015,Cuomo2019bis,GarmaOehmichen2019,Guo2019}). A different approach is based on the Fourier analysis of the images, which consists in the decomposition of the deprojected azimuthal luminosity profile of the galaxy into a Fourier series \citep{Aguerri2000}. Through this analysis, $R_{\rm bar}$  can be recovered from the luminosity contrasts between the bar and interbar intensity as a function of radial distance (e.g., \citealt{Debattista2002, Aguerri2003, Aguerri2015, Gerssen2003, Debattista2004, Corsini2007,Cuomo2019bis,Cuomo2019, Guo2019}) or studying the phase angle of the Fourier mode $m = 2$ (e.g., \citealt{Aguerri2003, Gerssen2003, Corsini2003, Corsini2007, Debattista2004}). Analysing the PA of the deprojected isophotal ellipses, $R_{\rm bar}$ is the position where the PA changes by a value of $10\degr$ from the PA of the ellipse with the maximum $\epsilon$ value (e.g., \citealt{Debattista2002, Aguerri2003, Corsini2003, Corsini2007,Cuomo2019}). Finally, it is possible to perform a photometric decomposition of the surface brightness distribution of the galaxy (e.g., \citealt{Aguerri2003, Corsini2003, Corsini2007, Gerssen2003,Cuomo2019}), which however depends on the adopted parametric laws for the different galaxy components. The choice of a single measurement method is usually limited by its own weaknesses, so usually more than one method is adopted to recover \rbar, which is then given by the combined results from the different applied methodologies (e.g., \citealt{Aguerri2003,Aguerri2015,Guo2019,Cuomo2019bis,Cuomo2019}). We collected the value of \rbar\ provided by previous works for each galaxy in the sample, which corresponds to the mean value obtained with the different methods adopted in the corresponding work. The values of $R_{\rm bar}$ for all the sample galaxies are listed in Tables~\ref{tab:sample_literature},~\ref{tab:sample_califa}, and~\ref{tab:sample_manga}.

\subsection{Bar strength}
\label{sec:sbar}

The value of $S_{\rm bar}$ describes the contribution of the bar to the galaxy potential and measures the non-axisymmetric forces produced by the bar. A variety of methods have been developed to measure $S_{\rm bar}$ and here, we briefly review those adopted for the sample galaxies. A first method is based on the intrinsic ellipticity of the bar, which can be obtained from the isophotal radial profile or from a photometric decomposition \citep[e.g.,][]{Cuomo2019}. A related approach consists in measuring the total non-axisymmetry strength parameter $Q_t$ \citep[e.g.,][]{Treuthardt2007}. Finally, $S_{\rm bar}$ can be recovered in various ways from a Fourier analysis, either from the peak of the ratio between the amplitudes of the $m = 2$ and $m = 0$ Fourier components (e.g., \citealt{Guo2019,Cuomo2019}) or from the integral of the ratio between the $m = 2$ and $m = 0$ Fourier components divided by \rbar\ \citep[e.g.,][]{Cuomo2019bis}. Since these different methods are connected to different bar properties, their results can considerably differ, even for the same object. 

In the sample of galaxies studied here, the strengths of the bars were not always measured in previous works; this is especially true for the literature subsample and for the galaxies in \cite{Aguerri2015} and \citet{GarmaOehmichen2019}. To have consistent measurements of $S_{\rm bar}$, we thus adopted the peak of the ratio between the amplitudes of the $m = 2$ and $m = 0$ Fourier components, when this was available from literature \citep{Kim2016vizier,Cuomo2019bis,Cuomo2019,Guo2019}, and derived \sbar\ for all the galaxies for which it was not already available and for which we have photometric data to perform the Fourier analysis. We obtained the uncertainties on \sbar\ by performing a Fourier analysis using the two halves of the deprojected azimuthal surface brightness, as in \citet{Cuomo2019bis}. Galaxies with missing values correspond to results taken from literature, in which uncertainties were not provided and we were not able to repeat the analysis. Three galaxies in the literature subsample are lacking an estimate of \sbar, since photometric data to perform the Fourier analysis are not available. The values of $S_{\rm bar}$ for all the sample galaxies are listed in Tables~\ref{tab:sample_literature},~\ref{tab:sample_califa}, and~\ref{tab:sample_manga}.

\subsection{Bar pattern speed}
\label{sec:pattern}

There are several methods to recover \omegabar\ (see \citealt{Corsini2011} for a discussion), but the only model-independent one is that based on the TW equation,

\begin{equation}
\langle V\rangle = \langle X\rangle \: \Omega_{\rm bar} \sin i,
\label{eq:tw}
\end{equation}

\noindent where $i$ is the disc inclination, and 

\begin{equation}
\langle X\rangle=\frac{\int X \Sigma dX}{\int \Sigma dX} ,\;\; \langle V\rangle=\frac{\int V_{\rm los}\Sigma dX}{\int \Sigma dX}
\label{eq.2}
\end{equation}

\noindent are the photometric, $\langle X\rangle$, and kinematic, $\langle V\rangle$, integrals, defined as the luminosity-weighted average of position $X$ and LOS velocity $V_{\rm los}$, respectively, where $\Sigma$ is the surface brightness, measured along directions parallel to the disc major axis and applied to a tracer population satisfying the continuity equation. In practice, the integrals are calculated along several apertures, one centred on the galaxy centre and the others with an offset, but all aligned with the disc major axis. The slope of the straight line defined by the measured values of $\langle X\rangle$ versus $\langle V\rangle$ gives $\Omega_{\rm bar} \sin i$.
When using integral-field spectroscopic data, the luminosity weights in the integrals are obtained by collapsing the datacube along the wavelength and the spatial directions of each pseudo-slit \citep{Cuomo2019}. As an alternative, the kinematic integrals can be directly obtained from the stellar velocity field using either a map of the surface brightness \citep{Aguerri2015,GarmaOehmichen2019,Guo2019} or stellar mass \citep{Aguerri2015} as a weight in the definition of the integral. However, the mass and light distributions often do not match well, particularly in the presence of ongoing star-formation, as is usually the case in late-type galaxies. \citet{Gerssen2007} explored with numerical simulations the effect of dust and star formation on the TW method, concluding it is negligible and its application can be extended to the full range of Hubble types. Moreover, \citet{Aguerri2015} showed that surface brightness and stellar mass used as weights for the integrals lead to consistent results.

Despite the simple formulation, the TW method has some critical issues as well. In fact, it assumes that a barred galaxy has a single, well-defined pattern speed. However, it can host components that are independently rotating, such as an inner bar and spiral arms \citep[e.g.,][]{Tagger1987,Sellwood1988,Sygnet1988}. \citet{Corsini2003} attempted the application of the method to the double-barred galaxy NGC~2950. They successfully derived the pattern speed of the primary bar and argued that the secondary one is independently rotating. Meanwhile, spiral arms perturb not just the TW measurements but also measurements of bar length \citep{Petersen2019,Hilmi2020}. To deal with such issues, methods taking into account the radial change of the pattern speed have been developed \citep[e.g.,][]{Meidt2008a,Meidt2008b}. The number of pseudo-slits used to measure the TW integrals is a critical element in the error budget of \omegabar\ \citep{Corsini2011}. Since \omegabar\ is related to the slope of a linear fit between the integrals, it is necessary to define at least three pseudoslits to get the slope (in principle two, but this is in practice not enough since both the integrals are affected by their own errors). Usually, the number of pseudo-slits is maximised according to the characteristics of the specific target and/or to the observations and its effect is taken into account in the uncertainty associated with \omegabar, as explored by \citet{Cuomo2019,Cuomo2020}.

The most critical parameter for the TW method is the correct definition of the PA of the disc along which to locate the pseudo-slits (\citealt[][]{Debattista2003,Cuomo2019,GarmaOehmichen2019,Zou2019}). 
Isophotal analysis, photometric decomposition, and kinemetry have been shown to give consistent results (\citealt{Aguerri2015, Guo2019, Cuomo2019bis}). 

To have consistent values of $\Omega_{\rm bar}$ for all the sample galaxies, we selected for our analysis the TW measurements adopting the photometric PA of the galaxy major axis, when more than one PA estimate was available (e.g., \citealt{Cuomo2019bis, GarmaOehmichen2019,Guo2019}), and the luminosity weight in the integrals. We collected the values of \omegabar\ provided by previous works for each galaxy in the sample. The values of $\Omega_{\rm bar}$ for all the sample galaxies are listed in Tables~\ref{tab:sample_literature},~\ref{tab:sample_califa}, and~\ref{tab:sample_manga}.

\subsection{Corotation radius and bar rotation rate}
\label{sec:rcr_r}

The value of \rcor\ is the radius where the gravitational and centrifugal forces cancel out in the rest frame of the bar; it is given by the ratio between $V_{\rm circ}$ and \omegabar. A simple and basic approach to compute $V_{\rm circ}$ consists in using the maximum of the (cold) gaseous rotational velocity as approximation for $V_{\rm circ}$ \citep[e.g.,][]{Treuthardt2007}. Another straightforward and more solid estimate consists in the application of the asymmetric drift correction to the observed stellar streaming velocities (e.g., \citealt{Merrifield1995, Debattista2002, Aguerri2003, Aguerri2015, Corsini2003, Corsini2007, Gerssen2003, Debattista2004,Cuomo2019bis,Cuomo2019}). A full dynamical model built from the kinematics and surface brightness of the stellar component represents a more sophisticated approach to recover \vcirc\ \citep[e.g.,][]{Gerssen1999,Guo2019}. Moreover, the value of \rcor\ can be directly estimated as the intersection between \omegabar\ and the modelled angular rotation curve \citep[e.g.,][]{GarmaOehmichen2019}. This approach is useful for galaxies where the rotation curve rises slowly and \rcor\ can be overestimated, when measured using \vcirc. 

The bar rotation rate \rr\ is given by the ratio between \rcor\ and \rbar\, and it is provided for all the galaxies in our sample. We collected the values of \rcor\ and \rr\ provided by previous works for each galaxy in the sample. The values of \rcor\ and the resulting ${\cal R}$ for all the sample galaxies are listed in Tables~\ref{tab:sample_literature},~\ref{tab:sample_califa}, and~\ref{tab:sample_manga}.

\subsection{Uncertainties on bar parameters}
\label{sec:bar_properties_uncertainties}

The successful application of the TW method requires the disc to have an intermediate $i$ and the bar to be located at an intermediate PA with respect to the disc major and minor axes. In fact, when the galaxy is too inclined, it is difficult to identify the bar and consequently to place the apertures. On the other hand, a low value of $i$ corresponds to low LOS velocities with large errors. Instead, if the bar is almost aligned to the disc major axis it is difficult to define a sufficient number of apertures. The other extreme case, when the bar is close to the disc minor axis, leads to low values of the photometric integrals. These extreme situations make it hard to apply the TW method and to control the errors, which translates into large uncertainties in the measured parameters \citep{Debattista2003, Corsini2011, Zou2019}. For these reasons, all the sample galaxies have an inclination $20\degr\lesssim i\lesssim 70\degr$ and a PA difference between bar and disc axes $10\degr \lesssim \Delta {\rm PA}\lesssim80\degr$. 

The disc inclination and the position of the bar with respect to the disc axes may also affect the correct estimation of \rbar\ \citep{Debattista2003, Corsini2011}. Moreover, recovering bar orientation and ellipticity from ellipse fits can be very difficult in real galaxies, especially when the galaxy is very inclined, as explored by \citet{Comeron2014}. Using toy models, the authors showed that orientation of the bar with respect to the line of nodes can be correctly recovered for intermediate inclination $i<60\degr$ when the bar is long and for $i<40\degr$ when the bar is small with respect to the bulge. For highly inclined galaxies, it is not always possible to obtain reliable bar properties.

\begin{figure*}[!t]
    \centering
    \includegraphics[scale=0.65]{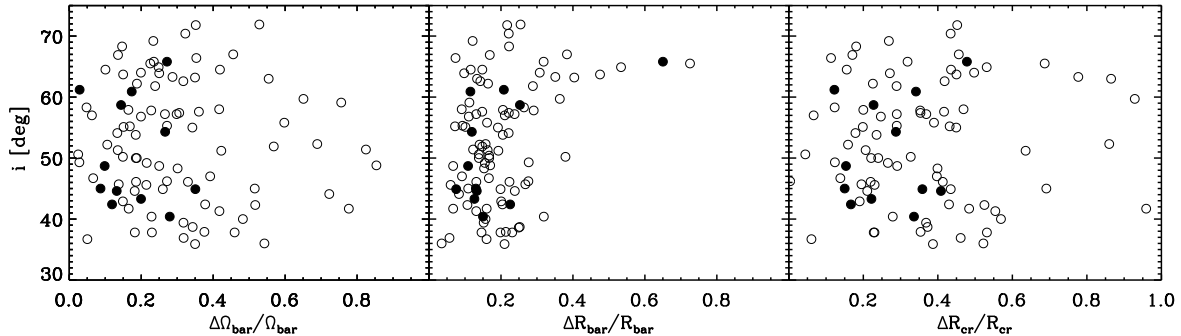}
    \caption{Relative errors of the bar pattern speed $\Delta \Omega_{\rm bar}/\Omega_{\rm bar}$, bar radius $\Delta R_{\rm bar}/R_{\rm bar}$, and corotation radius $\Delta R_{\rm cr}/R_{\rm cr}$ as a function of the disc inclination $i$. Ultrafast bars are highlighted ({\em black filled points}). Galaxies with relative errors larger than 100 per cent are not shown.}
    \label{fig:error_inc}
\end{figure*}
\begin{figure*}[!t]
    \centering
    \includegraphics[scale=0.65]{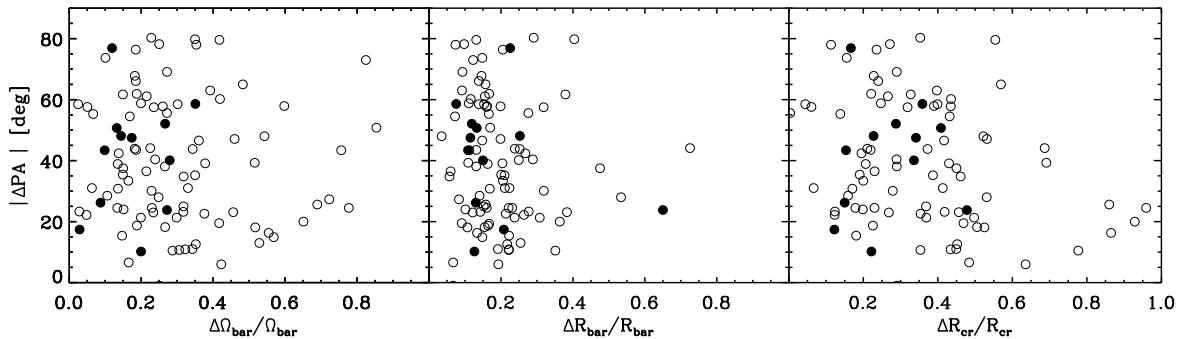}
    \caption{As in Fig.~\ref{fig:error_inc} but as a function of the bar orientation with respect to the disc major and minor axes $\Delta {\rm PA}$, given as the absolute value of the difference between the PA of the bar and that of the axes onto the sky plane.}
    \label{fig:error_deltaPA}
\end{figure*}
Figures~\ref{fig:error_inc} and \ref{fig:error_deltaPA} show the relative errors on \omegabar, \rbar, and \rcor\ for the sample galaxies as a function of the disc $i$ and bar $\Delta {\rm PA}$, respectively. Despite the large uncertainties of some values, we did not observe any significant trend and we excluded any selection bias on the galaxy sample. There is also no evident bias for the ultrafast bars. Indeed, 12 galaxies ($\sim12$ per cent) of the sample host a bar with ${\cal R}< 1.0$ at 95 per cent confidence level. 
All of them are found in LTBGs. Currently, we do not yet know whether ultrafast bars are an artifact of the TW method or a new class of objects that disprove the predictions of theory and numerical simulations about the size of the bar (see also \citealt{Aguerri2015} and \citealt{Guo2019}, for a discussion). This issue requires further investigation which is beyond the scope of this paper, and we therefore ignore the 12 ultrafast bars from the rest of our analysis.

\section{Results}
\label{sec:bar_properties_results}

\subsection{Results for the galaxy sample}
\label{sec:bar_properties_results_sample}

The mean relative errors and corresponding standard deviations in parenthesis on the bar parameters for the sample galaxies listed in Tables~\ref{tab:sample_literature},~\ref{tab:sample_califa}, and~\ref{tab:sample_manga} are $\langle\Delta \Omega_{\rm bar}/\Omega_{\rm bar}\rangle=0.35~(0.57)$, $\langle\Delta R_{\rm bar}/R_{\rm bar}\rangle=0.20~(0.11)$, and $\langle\Delta S_{\rm bar}/S_{\rm bar}\rangle=0.1~(0.1)$, leading to $\langle\Delta R_{\rm cr}/R_{\rm cr}\rangle=0.47~(0.56)$, and $\langle\Delta {\cal R}/{\cal R}\rangle=0.5~(0.4)$. In a number of galaxies, the uncertainties on \omegabar, \rcor, and \rr\ are larger than 100 per cent (this happens in 2 per cent of the sample galaxies for \omegabar, 9 per cent for \rcor, and 13 per cent for \rr). For a subsample of 53 galaxies we considered the bulge-to-total luminosity ratio $B/T$ from the photometric decompositions available in the literature ( \citealt{Treuthardt2007,Laurikainen2010,Salo2015,MendezAbreu2017,Kruk2018,Cuomo2019}, Tables~\ref{tab:sample_literature} and~\ref{tab:sample_califa}). In order to maximise and harmonise the number of galaxies with available $B/T$ values, we preferred bi-dimensional parametric decompositions, which include the possibility to model a bar and are based on SDSS $r$-band images or on other bands equivalently suitable to measure the bulge contribution. Indeed, \citet{Salo2015} used Spitzer $3.6\mu$m-band images, while \citet{Laurikainen2010} used $K_s$-band images. However, the adopted photometric decompositions do not usually include structural sub-components such as nuclear/inner lenses or barlenses. This was done in order to reduce the fit degeneracy \citep{MendezAbreu2017}, but this may affect the correct estimate of the bulge contribution \citep[see e.g.,][for a discussion]{Athanassoula2005,HerreraEndoqui2015,Laurikainen2017}.

After excluding the galaxies hosting an ultrafast bar, we considered a subsample with a trustworthy measurement of the bar pattern speed as only those with $\Delta \Omega_{\rm bar}/\Omega_{\rm bar}\leq0.5$. The final sample consists of 77 objects with 34 ETBGs (with Hubble stage ranging between ${\rm T}=-4$ to 1, i.e. SB0-SBa) and 43 LTBGs (with ${\rm T}$ between 2 and 7, i.e. SBab-SBc). They have mean relative errors and corresponding standard deviation in parenthesis of $\langle\Delta \Omega_{\rm bar}/\Omega_{\rm bar}\rangle=0.26~(0.12)$, $\langle\Delta R_{\rm bar}/R_{\rm bar}\rangle=0.20~(0.11)$, and $\langle\Delta S_{\rm bar}/S_{\rm bar}\rangle=0.09~(0.07)$ leading to $\langle\Delta R_{\rm cr}/R_{\rm cr}\rangle=0.40~(0.55)$ and $\langle\Delta {\cal R}/{\cal R}\rangle=0.4~(0.4)$. The subsample of galaxies with available $B/T$ ratio reduces to 42 objects. The mean values and the corresponding standard deviations in parenthesis of the bar parameters for the final selected sample, ETBGs, and LTBGs, are reported in Table~\ref{tab:mean_value}.

\begin{table}[]
    \centering
    \caption{Mean value and corresponding standard deviation in parenthesis} of the bar parameters for the final sample of 77 objects, for the 34 ETBGs, and for the 43 LTBGs.
    \begin{tabular}{lccc}
    \hline
    Parameter & All & ETBGs & LTBGs \\
    \hline
    $\langle \Omega_{\rm bar}\rangle$ [km s$^{-1}$ kpc$^{-1}$] & $48~(31)$ & $58~ (36)$ & $41~(24)$ \\
    $\langle R_{\rm bar}\rangle$ [kpc] & $5.2~(2.5)$ & $4.6~(2.3)$ & $5.7~(2.6)$ \\
    $\langle S_{\rm bar}\rangle$ & $0.5~(0.2)$ & $0.5~(0.1)$ & $0.5~(0.2)$ \\
    $\langle R_{\rm cr}\rangle$ [kpc] & $6.2~(4.6)$ & $5.9~(5.5)$ & $6.4~(3.8)$ \\
    $\langle {\cal{R}}\rangle$ & $1.3~(0.7)$ & $1.3~(0.9)$ & $1.2~(0.5)$ \\
    \hline
    \end{tabular}
    \label{tab:mean_value}
\end{table}

We investigated all the possible relations between the available parameters of the bars (\rbar, \sbar, \omegabar, \rcor , and \rr) and their host galaxies (Hubble type, $M_r$, and $B/T$) using the {\sc{idl}} task {\sc{r$\_$correlate}}~ \citep{Press1992b}, which computes the Spearman rank correlation $r$ of two populations and the corresponding two-sided significance $p$ of its deviation from the null hypothesis. The $p$ value measures how likely any observed correlation is only due to chance. Values close to 0 suggest that the correlation is strong and that the null hypothesis of no significant correlation is not correct. In this case, a large value of $p$ is expected even when the effect of the correlation on $r$ is less stringent. We estimated the number of standard deviations $\sigma$ by which the sum-squared difference of the ranks deviates from its null-hypothesis \citep{Press1992b}. The resulting correlation parameters are given in Table~\ref{tab:correlations}. We find: 

\begin{itemize}
 \item very strong correlations ($|r| \geq 0.7$ and $p < 0.01$, resulting in a 99 per cent confidence level that the correlation is not given by chance) between \omegabar\ and \rcor\ (bars with larger corotation radii rotate with lower pattern speeds), between \rbar\ and \rcor\ (longer bars have larger corotation radii), and between $M_r$ and \rbar\ (fainter galaxies host shorter bars); 
    
    \item strong correlations ($0.4 \leq |r| < 0.7$ and/or $0.01\leq p<0.05$, confidence level between 99 and 95 per cent) between \omegabar\ and \rbar\ (longer bars rotate with lower pattern speeds), between \rr\ and \omegabar\ or \rcor\ (fast bars rotate with higher pattern speeds and have smaller corotation radii), between $M_r$ and \omegabar\ or \rcor\ (fainter galaxies host bars which rotate with higher pattern speeds and have smaller corotation radii), and between \rbar\ and \sbar\ (shorter bars are weaker);
    
    \item weak correlations ($0.2 \leq |r| < 0.4$ and/or $0.05\leq p<0.1$, confidence level between 90 and 95 per cent) between \omegabar\ and Hubble type (bars in ETBGs rotate with higher pattern speeds), between \omegabar\ and \sbar\ (stronger bars rotate with lower pattern speeds), and between \rcor\ and \sbar\ (weaker bars have smaller corotation radii);
    
    \item no correlation ($|r| < 0.2$ and $p\geq0.1$) between Hubble type and $M_r$, \rbar, \sbar, \rcor, or \rr, between \rr\ and \rbar, \sbar, or $M_r$, between $M_r$ and \sbar, and between $B/T$ and \omegabar, \rbar, \sbar, \rcor, or \rr.
\end{itemize}

We checked and confirmed that the correlations remain mostly unchanged after splitting the final sample into ETBGs and LTBGs (Table~\ref{tab:correlations}). For ETBGs, the correlations \omegabar-\sbar\ and \sbar-\rcor\ become strong. In addition, a weak correlation \rr-\sbar\ appears and we found a strong $M_r$-\sbar\ correlation (brighter galaxies host stronger bars). For LTBGs, the correlation \rcor-\sbar\ becomes weak, a weak anti-correlation \rr-\sbar\ appears with respect to ETBGs, while the correlation \sbar-$M_r$ disappears.

Figures~\ref{fig:omega_relation} and \ref{fig:omega_relation_e_l} show the correlations found between \omegabar\ and Hubble type, $M_r$, \rbar, or \sbar\ before and after splitting the final sample between ETBGs and LTBGs, respectively. The points are colour-coded according to the value of \rr, highlighting the correlation between \rr\ and \omegabar. Figures~\ref{fig:rbar_relation} and \ref{fig:rbar_relation_e_l} show the correlations found between \rbar\ and $M_r$, \sbar, or \rcor\ before and after splitting the final sample between ETBGs and LTBGs, respectively. The points are colour-coded according to the nominal value of \rr, highlighting the correlation between \rr\ and \rcor. Figure~\ref{fig:bulge_nocut} shows the correlation between $B/T$ and \sbar\ for the subsample of 41 galaxies with an available photometric decomposition and an estimate of \sbar. The points are colour-coded according to \rr, highlighting the correlation \rr-\rcor\ and pointing that there is no correlation between \rr\ and $M_r$, \rbar, or \sbar.

\subsection{Results for the Milky Way}

\begin{figure*}[!t]
    \centering
    \includegraphics[scale=0.7]{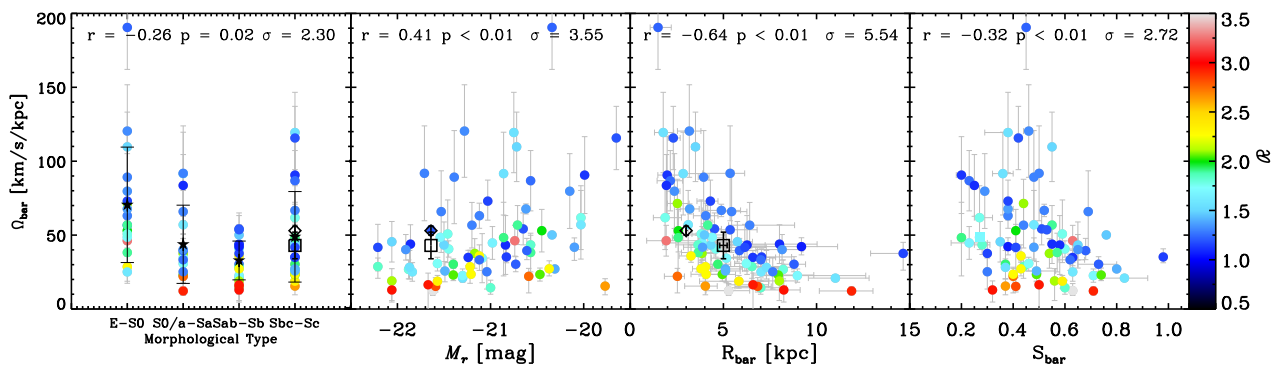}
    \caption{Relations between the bar pattern speed \omegabar\ and the Hubble type, SDSS $r$-band absolute magnitude $M_r$, bar radius \rbar, and bar strength \sbar\ for the selected sample of 77 galaxies. The Spearman rank correlation $r$, two-sided significance $p$, and number of $\sigma$ from the null-hypothesis are given in each panel. The points are colour-coded according to the value of \rr.
    Results obtained using different bands with respect to the SDSS $r$-band are represented with a square symbol. The mean value of \omegabar\ for each bin of Hubble type is shown ({\em black filled stars}). The results for the Milky Way are shown both for the short bar case ({\em black open diamond}) and long bar case ({\em black open square}).}
    \label{fig:omega_relation}
\end{figure*}

\begin{figure*}[!t]
    \centering
    \includegraphics[scale=0.7]{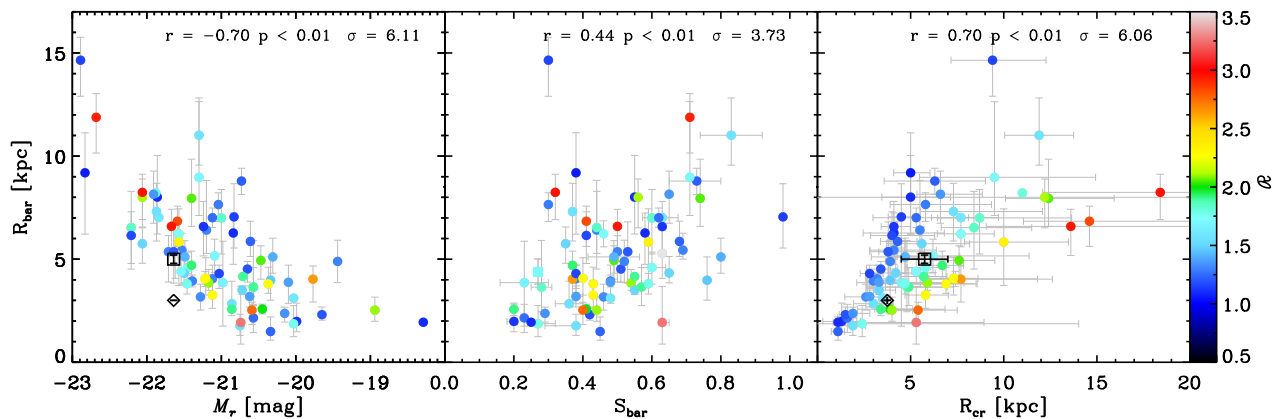}
    \caption{Relations between the bar radius \rbar\ and the SDSS $r$-band absolute magnitude $M_r$, bar strength \sbar, and corotation radius \rcor\ for the selected sample of 77 galaxies. The Spearman rank correlation $r$, two-sided significance $p$, and number of $\sigma$ from the null-hypothesis are given in each panel for the selected sample of 77 galaxies. The points are colour-coded according to the value of \rr. Results obtained using different bands with respect to the SDSS $r$-band are represented with a square symbol. The results for the Milky Way are shown both for the short bar case ({\em black open diamond}) and long bar case ({\em black open square}).}
    \label{fig:rbar_relation}
\end{figure*}

The Milky Way is a barred galaxy. Large effort has been made to derive the bar parameters of our own Galaxy and to understand whether it hosts a long or a short bar \citep[see][for a discussion]{BlandHawthorn2016}. The long bar case \citep{Portail2017,Donghia2020} implies that the Milky Way bar rotates with lower \omegabar\ with respect to the short bar case \citep{Dehnen2000,Fragkoudi2019}. These two alternative hypotheses give  different results for \rcor\, but similar values of \rr. We collected the available results for the long bar case from \cite{BlandHawthorn2016} and for the short bar case from \cite{Dehnen2000}. Our own galaxy has values of \rbar, \rcor, and \omegabar\ typical for a LTBG in both cases, but it lies slightly lower with respect to the relation \rbar-$M_r$ for the short bar case (Figs.~\ref{fig:omega_relation} and \ref{fig:rbar_relation}).

\subsection{Relations and sample selection}
The relations discussed in Sect.~\ref{sec:bar_properties_results_sample} are almost unaffected by the selection criteria used to define the final sample (i.e., no ultrafast bars and relative error on \omegabar\ smaller than 50 per cent). In fact, we calculated the Spearman correlation for each relation using the original 104 galaxy sample and we checked in particular the results observed between \omegabar\ and Hubble type, $M_r$, \rbar, and \sbar\ and between \rbar\ and $M_r$, \sbar, and \rcor. We verified they do not change significantly from the results discussed in Sect.~\ref{sec:bar_properties_results_sample}. The observed relations, including with \omegabar, remain (becoming even stronger for \omegabar-\sbar), except for the case of \omegabar\ and Hubble type. In fact, \omegabar\ becomes constant along the Hubble sequence with no clear decrease in the intermediate types. This could be explained by the fact that ultrafast bars, associated with ${\cal R}<1.0$ and partially implying a large value of \omegabar, are all found in galaxies later than SBb. In this case, then, the correlation disappears. On the other hand, the observed relations involving \rbar\ remain.

\begin{sidewaystable*}
\caption[Spearman parameters of the explored correlations]{\label{tab:correlations} Spearman parameters of the explored correlations within the properties of the galaxies.}
\begin{tabular}{ccccccccccccccccc}
\hline \hline
Correlation & Selected & sample & & & & ETBGs & & & & & LTBGs\\
\hline
 & $N$ &  $r$ & $p$ value & $\sigma$ && $N$ &  $r$ & $p$ value & $\sigma$ && $N$ &  $r$ & $p$ value & $\sigma$ \\
(1) & (2) & (3) & (4) & (5) && (2) & (3) & (4) & (5) && (2) & (3) & (4) & (5)  \\
\hline
(Hubble Type-\omegabar) & 77 & $-0.3$ & 0.02 & 2.3 && ... & ... & ... & ... && ... & ... & ... & ...\\
(Hubble Type-\rbar) & 77 & 0.2 & 0.2 & 1.3 && ... & ... & ... & ... && ... & ... & ... & ... \\
(Hubble Type-\rcor) & 75 & 0.1 & 0.4 & 0.8 && ... & ... & ... & ... && ... & ... & ... & ... \\
(Hubble Type-\rr) & 77 & $-0.06$ & 0.6 & 0.5 && ... & ... & ... & ... && ... & ... & ... & ... \\
(Hubble Type-\sbar) & 74 & $-0.1$ &  0.4 & 0.9 && ... & ... & ... & ... && ... & ... & ... & ... \\
(Hubble Type-$M_r$) & 77 & $-0.03$ & 0.8 & 0.2  && ... & ... & ... & ... && ... & ... & ... & ...\\

(\omegabar-\rbar) & 77 & $-0.6$ & $5\times10^{-10}$ & 5.5 && 34 & $-0.6$ & $ 7\times10^{-4}$ & 3.2 && 43 & $-0.6$ & $ 8\times10^{-5}$ &  3.7\\
(\omegabar-\rcor) & 77 & $-0.9$ & $ 6\times10^{-28}$ & 7.8 && 34 & $-0.9$ & $ 1\times10^{-12}$ & 5.1 && 43 & $-0.9$ & $ 3\times10^{-15}$ &  5.7\\
(\omegabar-\rr) & 77 & $-0.5$ & $ 5\times10^{-7}$ & 4.7 && 34 & $-0.5$ & $9\times10^{-4}$ & 3.1 && 43 & $-0.6$ & $ 2\times10^{-5}$ & 3.9\\
(\omegabar-\sbar) & 73 & $-0.3$ & $ 6\times10^{-3}$ & 2.7 && 32 & $-0.4$ & 0.01 & 2.4 && 42 & $-0.3$ & 0.06 & 1.9\\
(\omegabar-$M_r$) & 77 & 0.4 & $2\times10^{-4}$ & 3.6 && 34 & 0.4 & 0.03 & 2.2 && 43 & 0.4 & 0.02 & $2.3$\\

(\rbar-\rcor) & 77 & 0.7 & $ 2\times10^{-12}$ & $6.1$ && 34 & 0.7 & $1\times10^{-5}$ & 3.9 && 43 & 0.7 & $ 9\times10^{-7}$ &  4.4\\
(\rbar-\rr) & 77 & $ -0.08$ & 0.5 & 0.70 && 34 & $-0.09$ & 0.6 & 0.49 && 43 & $-0.1$ & 0.5 & 0.66\\
(\rbar-\sbar) & 74 & 0.4 & $1\times10^{-4}$ & $3.7$ && 32 & 0.5 & $ 5\times10^{-3}$ & 2.7 && 42 & $0.4$ & $ 4\times10^{-3}$ & $2.8$\\
(\rbar-$M_r$) & 77 & $-0.7$ & $ 1\times10^{-12}$ & 6.1 && 34 & $-0.7$ & $ 5\times10^{-7}$ & 4.3 && 43 & $-0.7$ & $ 1\times10^{-6}$ &  4.3\\

(\rcor-\rr) & 77 & 0.6 & $7\times10^{-9}$ & 5.2 && 34 & 0.6 & $ 2\times10^{-4}$ &  3.4 && 43 & $0.6$ & $ 9\times10^{-6}$ & $4.0$\\
(\rcor-\sbar) & 74 & 0.3 & $ 6\times10^{-3}$ & $2.7$ && 32 & 0.6 & $8\times10^{-4}$ & 3.1 && 42 & 0.2 & 0.3 & $1.1$\\
(\rcor-$M_r$) & 77 & $-0.6$ & $ 8\times10^{-8}$ & 4.9 && 34 & $-0.6$ & $ 3\times10^{-4}$ &  3.3 && 43 & $-0.6$ & $ 2\times10^{-4}$ &  3.5\\

(\rr-\sbar) & 74 &  0.03 & 0.8 & 0.22 && 32 & 0.3 & 0.08 & 1.7 && 42 & $-0.2$ & 0.3 & 1.1\\
(\rr-$M_r$) & 77 & $-0.1$ & 0.4 & 0.83 && 34 & $-0.1$ & 0.4 & 0.82 && 43 & $-0.08$ & 0.6 & 0.50\\

(\sbar-$M_r$) & 74 & $-0.2$ & 0.2 & 1.3 && 32 & $-0.5$ & $ 7\times10^{-3}$ & 2.6 && 42 & 0.003 & 1 & $0.020$\\

($B/T$-\omegabar) & 42 & 0.2 & 0.3 & 1.0  && ... & ... & ... & ... && ... & ... & ... & ...\\
($B/T$-\rbar) & 42 & $-0.03$ & 0.9 & 0.17  && ... & ... & ... & ... && ... & ... & ... & ...\\
($B/T$-\sbar) & 41 & 0.2 & 0.2 & 1.3  && ... & ... & ... & ... && ... & ... & ... & ...\\
($B/T$-\rcor) & 42 & $-0.01$ & 1.0 & 0.07  && ... & ... & ... & ... && ... & ... & ... & ...\\
($B/T$-\rr) & 42 & $ -0.02$ & 1.0 & 0.11 && ... & ... & ... & ... && ... & ... & ... & ...\\

\hline
(\rbar/\rpetro-Hubble Type) & 67 & $-0.3$ & 0.04 & 2.1 && ... & ... & ... & ... && ... & ... & ... & ...\\
(\rbar/\rpetro-\omegabar) & 67 & $-0.2$ & 0.2 & 1.8  && ... & ... & ... & ... && ... & ... & ... & ...\\
(\rbar/\rpetro-\sbar) & 67 & 0.5 & $1\times10^{-5}$ & 4.1 && ... & ... & ... & ... && ... & ... & ... & ...\\
(\rbar/\rpetro-$M_r$) & 67 & 0.2 & 0.2 & 1.3 && ... & ... & ... & ... && ... & ... & ... & ...\\
(\rbar/\rpetro-\rr) & 67 & $-0.09$ & 0.5 & 0.74 && ... & ... & ... & ... && ... & ... & ... & ...\\
(\rbar-\rpetro) & 67 & 0.7 & $1\times10^{-10}$ & 5.6 && ... & ... & ... & ... && ... & ... & ... & ...\\
(\rcor/\rpetro-Hubble Type) & 67 & $-0.2$ & 0.07 & 1.8 && ... & ... & ... & ... && ... & ... & ... & ...\\
(\rcor/\rpetro-\omegabar) & 67 & $-0.6$ & $5\times10^{-7}$ & 4.6 && ... & ... & ... & ... && ... & ... & ... & ...\\
(\rcor/\rpetro-\sbar) & 67 & 0.4 & $1\times10^{-3}$ & 3.1 && ... & ... & ... & ... && ... & ... & ... & ...\\
(\rcor/\rpetro-$M_r$) & 67 & $-0.02$ & 0.9 & 0.17 && ... & ... & ... & ... && ... & ... & ... & ...\\ 
(\rcor/\rpetro-\rr) & 67 & 0.7 & $2\times10^{-11}$ & 5.7 && ... & ... & ... & ... && ... & ... & ... & ...\\
(\rcor-\rpetro) & 67 & 0.5 & $2\times10^{-6}$ & 4.4 && ... & ... & ... & ... && ... & ... & ... & ...\\ 
(\rbar/\rpetro-\rcor/\rpetro) & 67 & 0.6 & $1\times10^{-7}$ & 4.8 && ... & ... & ... & ... && ... & ... & ... & ...\\
\hline
\end{tabular}
\tablefoot{(1) Correlated parameters. (2) Number of galaxies used to explore the correlation. (3) Spearman rank correlation $r$ parameter. (4) Two-sided significant $p$ value of the correlation. (5) Number of standard deviation $\sigma$ from null-hypothesis.}
\end{sidewaystable*}

\begin{figure}
    \centering
    \includegraphics[scale=0.68]{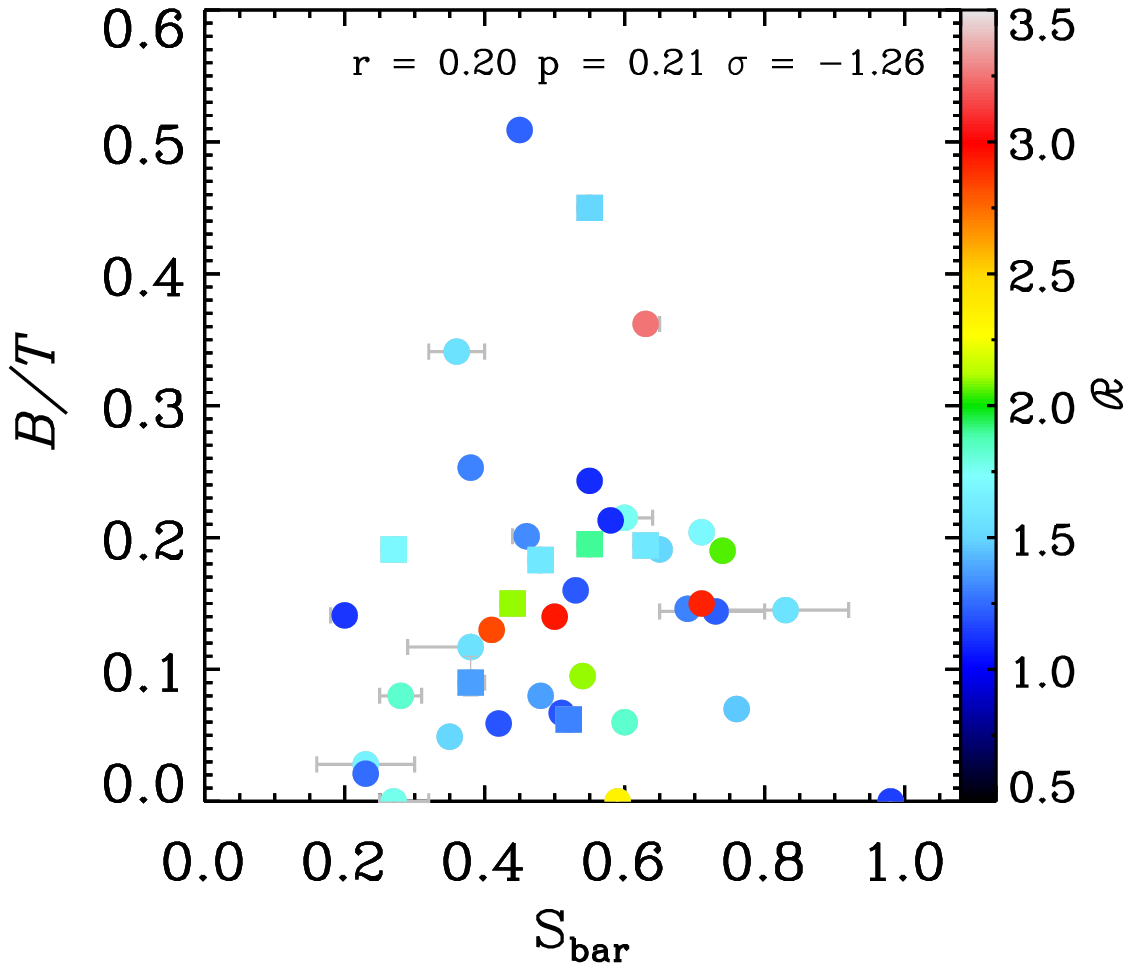}
    \caption[Relation between the bulge-to-total luminosity ratio and bar strength]{Relation between the bulge-to-total luminosity ratio $B/T$ and bar strength \sbar\ for the subsample of 41 galaxies with a photometric decomposition. The Spearman rank correlation $r$, the two-sided significance $p$, and number of $\sigma$ from the null-hypothesis are given. The points are colour-coded according to the value of \rr. Results obtained using different bands with respect to the SDSS $r$-band are represented with a square symbol.}
    \label{fig:bulge_nocut}
\end{figure}

\section{Discussion}
\label{sec:bar_properties_discussion}

\subsection{Relations among the bar parameters}

Some of the relations we reported among the bar parameters of TW-measured galaxies either confirm earlier observational findings and/or theoretical predictions (\omegabar-\rbar, \omegabar-\sbar, \omegabar-$M_r$, \rbar-\sbar, \sbar-\rcor, \rbar-$M_r$, \rr-\omegabar, and \omegabar-\rcor; Figs.~\ref{fig:omega_relation} and~\ref{fig:rbar_relation}). Others (\rbar-\rcor; Fig.~\ref{fig:rbar_relation}) are related to fact that the TW method finds mainly fast bars. Some bar parameters are not correlated at all (\rr-\rbar\ and \rr-\sbar; Figs.~\ref{fig:omega_relation} and~\ref{fig:rbar_relation}) because of the large scatter in their distribution. For example, the weak \rr-\sbar\ relation found for ETBGs is found to be an anti-correlation for LTBGs, giving no correlation in the final sample. Moreover, we did not find any or very weak relations between Hubble type and bar properties (Hubble type-\rbar, Hubble type-\sbar, Hubble type-\rcor, and Hubble type-\rr), even if some were previously suggested. The only exception is the weak Hubble type-\omegabar\ correlation, which however disappears when considering the original sample of 104 galaxies (Fig.~\ref{fig:omega_relation}). Moreover,
a large fraction of the galaxies in our sample come from \citet{Guo2019}, who reported some of the relations discussed here.

It was already known that longer bars are stronger (\citealt{Erwin2005,Kruk2018,Guo2019,Cuomo2019bis}) (with a tighter correlation among ETBGs, as reported by \citealt{DiazGarcia2016b}), but we found that they are also rotating with a lower \omegabar\ (Fig.~\ref{fig:omega_relation}, right panel). Such a \omegabar-\sbar\ relation was theoretically predicted \citep{Sellwood1981,Debattista2000,athanassoula2003,villavargas2010,Athanassoula2013}. It can be explained in terms of bar evolution because of the exchange of angular momentum during bar evolution, which can vary according to different properties of the host galaxy, such as the initial disc hotness/coldness, as investigated by \cite{athanassoula2003}. Our observational results are in agreement with the predictions of these numerical simulations. Although we are not able to probe the time evolution of a single bar, we conclude that the measured values of \omegabar\ and \sbar\ shown in the right panel of Fig.~\ref{fig:omega_relation} represent snapshots of bars in galaxies with different properties. The relation \omegabar-\sbar\ was already explored in \citet{Cuomo2019bis}, where we did not find any difference between the distributions of \omegabar\ in strongly and weakly barred galaxies defined according to \sbar. This was probably due to the limited number of sample galaxies, and/or because weak bars are expected to rotate with both high and low \omegabar, depending on other galaxy parameters. \citet{Font2017} discussed the correlation by plotting \omegabar\ scaled by the disc angular velocity versus \sbar\ and colour-coding the data according to the relative \rbar\ (their Fig. 9). Their largest bars appear to rotate more slowly, whereas the shortest bars have higher relative angular rates. They also observed that the strongest bars rotate with the lowest values of scaled \omegabar, while the bars that rotate with low scaled \omegabar\ can only have lower values of \sbar. However, they cautiously concluded that their results can not probe any trends due to bar evolution.

Moreover, the bulge is expected to play an important role in the exchange of angular momentum within a barred galaxy. Simulations showed that a bulge can gain angular momentum from the bar, especially from the inner part of the bar, at the Inner Lindblad Resonance \citep{athanassoula2003,Saha2016,Kataria2019}. We explored the possible role of the bulge in the evolution of barred galaxies but we did not find any correlations between $B/T$ and \sbar, \rbar\ or \omegabar, which could have confirmed the bulges act as efficient sinks of angular momentum from bars (Fig.~\ref{fig:bulge_nocut}). In fact, other factors influence the efficiency of the angular momentum exchange. \citet{athanassoula2003} showed that dynamically hot absorbers hamper its transfer, in agreement with the analytical results by \citet{LyndenBell1972} and \citet{Tremaine1984}. \citet{Combes1993} suggested that \omegabar\ critically depends on the relative bulge mass and $h$: galaxies with a large bulge-to-disc mass ratio tends to form rapidly rotating bars. These considerations suggest that many bulges in the analysed sample do not act as classical bulges. Moreover, there are three bulgeless galaxies ($B/T=0.0$) in our sample, with a wide range of \sbar\ (Fig.~\ref{fig:bulge_nocut}). A weak $B/T$-\sbar\ correlation appears when they are discarded from the analysis ($r=0.3$, $p=0.09$, $\sigma=1.7$). This result requires a further investigation on the behaviour of bulgeless barred galaxies. 


The significant relations found among \omegabar\ and \rbar\ and other galaxies properties are highlighted in Figs.~\ref{fig:omega_relation} and ~\ref{fig:rbar_relation}. The points are colour-coded according to the value of \rr. This allows to highlight the relations involving such a third parameter. Figure~\ref{fig:omega_relation} shows the \omegabar-\rr\ relation in each panel, while the \rr-\rcor\ relation is visible in the right-panel of Fig.~\ref{fig:rbar_relation}. More interestingly, no relation is found either between \rr\ and \rbar\ even if they are bound by definition, or between \rr\ and \sbar\ despite the strong \rbar-\sbar\ relation. We argue that the observed relations are driven by galaxy evolution: a bar evolves by shedding angular
momentum to material at resonance, making the bar slow down and in the process get longer and stronger. As the bar slows down,
\rcor\ is pushed outwards, where the density of the stars which can be trapped by the bar decreases. At some point, the density
becomes too low, so the bar size will not keep pace with \rcor. The relations \omegabar-\sbar, \sbar-\rbar, \omegabar-\rbar, and \rcor-\rbar\ fit well with the regime where the bar keeps pace with the slowdown. However, all the selected bars analysed so far are compatible with this fast regime at 95 per cent confidence level (despite some of them presenting nominal values of \rr\ larger than 1.4, we cannot exclude that these bars are fast taking into account the corresponding error on \rr): this explains why no relations \rr-\rbar\ and \rr-\sbar\ are observed. In the context of galaxy evolution, this could be explained if the sample does not include either dynamically evolved barred galaxies or cases in which the exchange of angular momentum is very efficient. The observed lack of slow bars requires further investigation, both from a theoretical and an observational point of view.

One could argue that the resulting relations are mostly driven by galaxy total mass rather than secular evolution. In particular, the \omegabar-$M_r$ relation could be seen as a relation between the pattern speed and galaxy total mass through the well-known \vcirc-$M_r$ relation \citep{Tully1977}. We explored the possible role of the galaxy total mass and claim it is not the main driver of the resulting relations. In fact, we did not observe any correlation between \omegabar\ and \vcirc\ in our sample of galaxies. Moreover, when discarding galaxies brighter than $-21$ mag from our final sample, the relations among the bar properties hold even if we do not observe a \omegabar-$M_r$ relation. In this case the Spearman rank correlations and significances for the subsample of 33 galaxies become $r=0.1$ and $p=0.4$ for the \omegabar-$M_r$ relation, whereas we got $r=0.4$ and $p=0.03$ for \omegabar-\sbar\ relation, $r=0.6$ and $p=8\times10^{-5}$ for \sbar-\rbar\ relation, $r=-0.7$ and $p=2\times10^{-5}$ for \omegabar-\rbar\ relation, and $r=0.6$ and $p=7\times10^{-5}$ for \rcor-\rbar\ relation. However, it could be more significant to explore the effects of the \omegabar-$M_r$ relation by defining several bins in magnitude. Unfortunately, this is not feasible here due to the small sample size. Upcoming observations for large samples of galaxies will help to further explore this issue (e.g., \citealt{Balcells2010,Croom2012,bundy2015}).

\subsection{Relations with the galaxy luminosity}

We observed interesting relations between the bar parameters and galaxy luminosity. In particular, brighter galaxies host longer bars, which rotate with lower \omegabar\ and have larger corotation radii (Figs.~\ref{fig:omega_relation} and~\ref{fig:rbar_relation}). We verified that these strong correlations were not driven by fainter/brighter galaxies and/or by galaxies with bars rotating with very high \omegabar. The relations hold even when we split the final sample between ETBGs and LTBGs (Fig.~\ref{fig:omega_relation_e_l} and~\ref{fig:rbar_relation_e_l}). 

\cite{sheth2008} studied the bar fraction over $0.2<z<0.84$ with a sample of more than 2000 luminous face-on spirals from the COSMOS survey \citep{Scoville2007}. The presence of a bar strongly correlates with both the stellar mass and bulge prominence. In fact, the bar fraction in very massive and luminous spirals ($M_\ast>10^{10.9}\msun$, $M_V<-23.5$ mag) is roughly constant out to $z\sim0.84$, whereas for the low-mass blue spirals ($M_\ast<10^{10.5}\msun$, $M_V>-22.5$ mag) it significantly declines beyond $z=0.3$. On the other hand, the bar fraction at low redshift is roughly equal at all luminosities. The bar fraction at high redshift turned out to be slightly higher for bulge-dominated galaxies, suggesting a co-evolution of bars and bulges. At low redshift, this trend disappears and the bar fraction is roughly constant for all Hubble types, although only a few bulgeless galaxies are observed. \cite{sheth2008} concluded that their results are a clue for a downsizing process in the formation of bars: the more massive and luminous galaxies have a higher bar fraction at higher redshift, which is close to the present-day value, whereas the less massive and luminous systems formed the majority of their bars at $z<0.8$. The early presence of bars in massive galaxies suggests that these systems became dynamically cool and sufficiently massive to host a bar at earlier times. On the contrary, the less massive systems have a low bar fraction because they are either dynamically hot, or not rotationally supported, or have not accreted sufficient mass to form a bar at high redshift.

Our results support this scenario: we found in brighter galaxies smaller \omegabar\ values together with larger \rbar\ and \rcor\ values, which are a signature of bar evolution in agreement with the idea these bars may have formed earlier and had sufficient time to slow down, grow in length, and push corotation outwards. 

The conclusions from \citet{sheth2008} has been subsequently partially questioned. In particular, \cite{erwin2018,Erwin2019} argued that those may be affected by a detection problem. Indeed, SDSS-based studies preferentially miss bars in lower-mass, bluer, and gas-rich galaxies due to poor spatial resolution, whereas he found bars as common in blue, gas-rich galaxies as they are in red, gas-poor galaxies using Spitzer data \citep{erwin2018}. The absence of any dependence of the presence of the bar and its size or the gas mass fraction has brought \cite{Erwin2019} to question theoretical models in which bar formation is suppressed by the high gas fraction in the disc. Moreover, \cite{GarmaOehmichen2019} measured the molecular gas fraction in their sample of 18 galaxies and found a significant correlation with \rr\ and a weak anti-correlation with \omegabar. All the galaxies with a small gas fraction are consistent with the fast - ultrafast regime, leading the authors to conclude this could be an indication that bars do slow down more efficiently in gaseous discs.

Moreover, the claim from \citet{sheth2008} of a strong correlation between the presence of a bar and bulge prominence has been disputed by recent work, which highlighted a large fraction of barred galaxies in late-type galaxies ($T\geq5$; \citealt{Buta2015,DiazGarcia2019}), many of which are known to be bulgeless \citep{Salo2015} or to host very small bulges \citep{Costantin2020}. Our results support the idea that a possible co-evolution between bulge and bar is not confirmed by the absence of correlations between $B/T$ and the main bar parameters. Only a weak $B/T$-\sbar\ correlation appears when excluding the bulgeless galaxies from our sample (Fig.~\ref{fig:bulge_nocut}).

\subsection{Relations involving the normalised sizes of galaxies}

We collected the Petrosian radius \rpetro\ and corresponding error provided by the SDSS in the $r$-band for our original sample of galaxies when available. According to the selection criteria described in Sect.~\ref{sec:rcr_r} and \ref{sec:bar_properties_results} and the availability of \rpetro, we obtain a restricted sample of 67 galaxies. Then we normalised \rbar\ and \rcor\ by \rpetro\ and we reanalysed all the correlations listed in Table~\ref{tab:correlations}, measuring the Spearman correlation parameters. We observe a very strong correlation between \rbar\ and \rpetro\ (larger bars correspond to larger Petrosian radii); a strong correlation between scaled \rbar\ and scaled \rcor\ (shorter bars according to the normalised size are associated with shorter normalised corotation radii), between \sbar\ and normalised \rbar\ and \rcor\ (weaker bars are shorter according to the normalised size and have shorter normalised corotation radii), between \omegabar\ and normalised \rcor\ (bars rotating with large pattern speeds present shorter normalised corotation radii), and between \rcor\ and \rpetro\ (larger corotation radii correspond to larger Petrosian radii); a weak correlation between Hubble type and both normalised \rbar\ and \rcor\ (late type galaxies have shorter normalised bars and shorter normalised corotation radii); no correlation between $M_r$ and normalised \rbar\ or normalised \rcor\, and between \omegabar\ and normalised \rbar\ (Fig.~\ref{fig:scaled_relation}). When considering the normalised sizes, the main difference with respect to previous discussed results is that the correlations $M_r$-normalised \rbar\ and $M_r$-normalised \rcor\ and \omegabar-normalised \rbar\ disappear. However, a weak \omegabar-normalised \rbar\ correlation appears ($r=0.2$, $p=0.08$, $\sigma=1.8$) when discarding the point corresponding to the double-barred galaxy NGC~2950 with large \omegabar\ and large normalised \rbar. A strong \sbar-\rbar\ correlation after normalising bar lengths to the disc size was previously reported by \cite{Elmegreen2007} and \citet{Gadotti2011}.

\section{Conclusions}
\label{sec:conclusions}

We took into account all the barred galaxies available in the literature with a direct measurement of \omegabar\ obtained with the TW method from long-slit or IFU spectroscopic data of stellar kinematics. We gathered the data from \citet{Corsini2011}, \citet[][]{Cuomo2019bis,Cuomo2019}, \citet{Aguerri2015}, \cite{GarmaOehmichen2019}, and \citet{Guo2019}. The sample consists of 104 galaxies with Hubble types ranging from SB0 to SBd, redshifts $z<0.08$, and SDSS $r$-band absolute magnitudes $-23 < M_r < -18$ mag. For each galaxy, we collected the values of the bar radius \rbar, bar strength \sbar, bar battern speed \omegabar, corotation radius \rcor, and bar rotation rate \rr. To have consistent measurements of \sbar, we derived it from the $m=2$ normalised bar Fourier amplitude following \cite{Athanassoula2002} for galaxies for which it was not already available in literature. We also collected the $B/T$ ratio for a subsample of galaxies with an available photometric decomposition.

The successful application of the TW method requires the disc to have an intermediate inclination $i$ and the bar to be located at an intermediate PA with respect to the disc major and minor axes. We checked that these selection criteria do not systematically affect the uncertainties on \rbar, \omegabar, and \rcor. Moreover, there is no bias for the ultrafast bars (${\cal R}< 1.0$) we found in 12 LTBGs of the sample ($\sim12$ per cent). At the moment, we do not yet know whether ultrafast bars are an artifact of the TW method or a new class of objects that violates the predictions of theory and numerical simulations about the extension of the bar, even if the fact that they are found only in late-type galaxies may strengthen the first hypothesis \citep[but see also][for a discussion]{Aguerri2015, Guo2019}. This issue requires further investigation, possibly based on dynamical modelling \citep[e.g.,][]{Vasiliev2020}, but this is beyond the scope of the paper. Therefore, we decided to consider only the 77 sample galaxies with a small relative error on \omegabar\ ($\Delta \Omega_{\rm bar}/\Omega_{\rm bar}\leq0.5$) and not hosting an ultrafast bar (${\cal R}<1$). We investigated all the possible relations between the available bar parameters (\rbar, \sbar, \omegabar, \rcor , and \rr) and galaxy properties (Hubble type, $M_r$, and $B/T$) and discussed their significance. Some of the relations we reported confirm earlier observational findings and/or theoretical predictions (\omegabar-\rbar, \omegabar-\sbar, \omegabar-$M_r$, \rbar-\sbar, \sbar-\rcor, \rbar-$M_r$, \rr-\omegabar, and \omegabar-\rcor). We verified that the stronger relations are not driven by fainter/brighter galaxies, galaxies with very fast bars, and Hubble type. Moreover, the correlation \omegabar-$M_r$ is not driving the other ones. In fact, it is possible to define a luminosity cut subsample for which the correlation \omegabar-$M_r$ does not hold, while all the others do, and no \omegabar-\vcirc\ correlation is found.

In particular, we found that stronger bars rotate slower. Such a \omegabar-\sbar\ relation was theoretically predicted but never clearly observed. It can be explained in terms of bar evolution because of the exchange of angular momentum during bar evolution depending on galaxy properties, as numerically investigated by \cite{athanassoula2003}. We also reported that brighter galaxies host longer bars, which rotate slower and have a larger corotation. This observational finding is in agreement with a scenario of downsizing in bar formation if more massive galaxies formed earlier and had sufficient time to slow down, grow in length, and push outwards corotation \citep{sheth2008}. The possible role of the bulge and its co-evolution with the bar remain unclear.

Both the predictions of a long and short bar in the Milky Way are in agreement with the findings obtained from our sample. In fact, the values obtained for \omegabar, \rbar, \rcor, and $M_r$ for both the models end up in the region of the explored relations spanned by our sample (i.e., of the \omegabar-\rbar, \omegabar-$M_r$, \rbar-\rcor\ relations; Figs.~\ref{fig:omega_relation} and ~\ref{fig:rbar_relation}). However, the fact that the short bar model is slightly below the explored \rbar-$M_r$ relation favours the long bar.

\section*{Acknowledgements} 
We are grateful to the anonymous referee for the detailed comments that helped us to improve this paper. We thank M. Bureau and F. Combes for their insightful comments and C. Buttitta for her help in the revision of the paper. VC acknowledges support by Fondazione Ing. Aldo Gini and thanks the Instituto de Astrofísica de Canarias, Universidad de la Laguna, and University of Central Lancashire for hospitality during the preparation of this paper. VC and EMC acknowledge support by Padua University through grants DOR1715817/17, DOR1885254/18, and DOR1935272/19. EMC is also supported by MIUR grant PRIN 2017 20173ML3WW\_001. JALA is supported by the Spanish MINECO grant AYA2013-43188-P. VPD is supported by STFC Consolidated grant ST/R000786/1.




\bibliographystyle{aa}



\begin{appendix}

\section{Properties of galaxies}
\label{appendix:a}

We tabulated here all the parameters used in this work collected for our parent sample of 104 galaxies. 

\begin{sidewaystable*}
\small
\caption[Properties of the literature subsample]{\label{tab:sample_literature} Properties of the 18 galaxies of the literature subsample.}
\begin{center}
\begin{tabular}{cccccccccccc}
\hline\hline
Galaxy & Morph. Type & $z$ & $M_r$ & $R_{\rm bar}$ & $S_{\rm bar}$ & \omegabar\ & \rcor\ & \rr\ & $B/T$ (Ref.) & Ref. & Final Sample\\
 & & & [mag] & [kpc] &  & [km s$^{-1}$ kpc$^{-1}$] & [kpc] &  &  & \\
 (1) & (2) & (3) & (4) & (5) & (6) & (7) & (8) & (9) & (10) & (11) & (12)\\ 
\hline
ESO 139-G09 & SAB0$^0$(rs) & 0.018 & $-21.20$ & 6.41$^{+2.41}_{-1.13}$ & 0.44 & $56.7\pm15.4$ & 5.54$^{+1.96}_{-1.17}$ & 0.8$^{+0.3}_{-0.2}$ & ... & 1 & yes\\
ESO 281-G31 & SB0$^0$(rs) & 0.018 & $-21.22$ & $4.04\pm0.37$ & ... & $28.6\pm11.2$ & 7.35$^{+4.41}_{-1.47}$ & 1.8$^{+1.1}_{-0.4}$ & ... & 2 & yes\\
IC 874 & SB0$^0$(rs) & 0.008 & $-20.57$ & 3.65$^{+0.94}_{-0.90}$ & 0.57 & $38.2\pm13.1$ & 4.93$^{+2.40}_{-1.25}$ & 1.4$^{+0.7}_{-0.4}$ & ... & 1 & yes\\
NGC 271 & (R’)SBab(rs) & 0.014 & $-21.81$ & $7.70\pm0.27$ & 0.63 & $29.5\pm16.0$ & 11.69$^{+7.97}_{-4.25}$ & 1.5$^{+1.0}_{-0.5}$ & ... & 2 & no\\
NGC 936 & SB0$^+$(rs) & 0.005 & $-20.71$ & 4.16 & 0.55 & $56.8\pm13.2$ & 5.74$^{+1.25}_{-1.25}$ & 1.4$^{+0.5}_{-0.4}$ & 0.20 (1) & 3 & yes\\
NGC 1023 & SB0$^-$(rs) & 0.002 & $-21.39$ & 3.94$^{+0.29}_{-0.29}$ & 0.48 & $89.2\pm31.5$ & 3.03$^{+0.34}_{-0.34}$ & 0.8$^{+0.5}_{-0.3}$ & ... & 4 & yes\\
NGC 1308 & SB0/a(r) & 0.021 & $-21.71$ & 5.36$^{+0.78}_{-1.47}$ & 0.50 & $91.8\pm32.1$ & 3.85$^{+1.95}_{-0.99}$ & 0.8$^{+0.4}_{-0.2}$ & ... & 1 & yes\\
NGC 1358 & SAB0/a(r) & 0.013 & $-21.09$ & $5.16\pm0.81$ & ... & $34.4\pm16.6$ & 6.24$^{+5.16}_{-1.90}$ & 1.2$^{+1.0}_{-0.4}$ & ... & 2 &  yes\\
NGC 1440 & (R')SB0$^0$ & 0.005 & $-18.94$ & 2.53$^{+0.63}_{-0.54}$ & 0.44 & $71.4\pm16.4$ & 3.97$^{+1.11}_{-0.73}$ & 1.6$^{+0.5}_{-0.3}$ & 0.15 (2) & 1 & yes\\
NGC 2523 & SBbc(r) & 0.012 & $-21.88$ & 8.21 & ... & $26.8\pm6.4$ & 10.99 & 1.3$^{+0.7}_{-0.5}$ & 0.07 (3) & 5 & yes\\
NGC 2950 & (R)SB0$^0$(r) & 0.004 & $-20.72$ & $3.50\pm0.21$ & 0.55 & $109.7\pm23.5$ & 3.30$^{+0.89}_{-0.63}$ & 1.0$^{+0.3}_{-0.2}$ & 0.45 (2) & 6 & yes\\
NGC 3412 & SB00(s) & 0.003 & $-20.45$ & $2.58\pm0.24$ & 0.41 & $53.0\pm14.4$ & 3.90$^{+1.44}_{-0.82}$ & 1.5$^{+0.6}_{-0.3}$ & ... & 1 & yes\\
NGC 3992 & SBbc(rs) & 0.004 & $-19.44$ & $4.89\pm1.03$ & 0.52 & $66.7\pm4.2$ & $3.86\pm0.26$ & 0.8$^{+0.2}_{-0.2}$ & 0.06 (1) & 2 & yes\\
NGC 4245 & SB0/a & 0.003 & $-20.03$ & 3.11 & 0.48 & $57.0\pm23.2$ & 3.49 & 1.1$^{+1.1}_{-0.4}$ & 0.18 (1) & 5 & yes\\
NGC 4264 & SB0 & 0.008 & $-20.62$ & $3.19\pm0.51$ & 0.38$^{+0.02}_{-0.01}$ & $67.7\pm3.4$ & $2.81\pm0.17$ & 0.88$^{+0.23}_{-0.23}$ & 0.09 (4) & 7 & yes\\
NGC 4431 & dSB0/a & 0.003 & $-18.29$ & $1.94\pm0.13$ & 0.25 & $83.6\pm20.9$ & 1.12$^{+0.38}_{-0.26}$ & 0.6$^{+1.2}_{-0.4}$ & ... & 8 & yes\\
NGC 4596 & SB0$^+$(r) & 0.006 & $-21.84$ & 7.02 & 0.63 & $25.1\pm6.3$ & 7.72$^{+2.02}_{-1.58}$ &  1.1$^{+0.7}_{-0.3}$ & 0.19 (1) & 9 & yes\\
NGC 7079 & SB0$^0$(s) & 0.009 & $-21.54$ & $4.40\pm0.61$ & ... & $48.5\pm1.2$ & 5.30$^{+0.24}_{-0.23}$ & 1.2$^{+0.3}_{-0.2}$ & 0.19 (1) & 10 & yes\\
\hline
\end{tabular}\end{center}
\tablefoot{(1) Galaxy name. (2) Morphological classification from \cite{Corsini2011}, except for NGC~4264 \citep{Cuomo2019}. (3) Redshift from NED. (4) Absolute SDSS $r$-band magnitude obtained as described in Sect.~\ref{sec:sample}. (5) Bar radius obtained as described in Sect.~\ref{sec:abar}. (6) Bar strength obtained as described in Sect.~\ref{sec:sbar}. (7) Bar pattern speed obtained as described in Sect.~\ref{sec:pattern}. (8) Bar corotation radius obtained as described in Sect.~\ref{sec:rcr_r}. (9) Bar rotation rate obtained as described in Sect.~\ref{sec:rcr_r}. (10) Bulge-to-total luminosity ratio provided by  1 = \cite{Salo2015}, 2 = \cite{Laurikainen2010}, 3 = \cite{Treuthardt2007}, 4 = \cite{Cuomo2019} (11) Reference paper for the direct measurement of the bar pattern speed: 1 = \cite{Aguerri2003}, 2 = \cite{Gerssen2003}, 3 = \cite{Merrifield1995}, 4 = \cite{Debattista2002}, 5 = \cite{Treuthardt2007}, 6 = \cite{Corsini2003}, 7 = \cite{Cuomo2019}, 8 = \cite{Corsini2007}, 9 = \cite{Gerssen1999}, 10 = \cite{Debattista2004}. (12) Inclusion in the final analysed sample (galaxies hosting an ultrafast bar at 95 per cent confidence level and with $\Delta \Omega_{\rm bar}/\Omega_{\rm bar}>0.5$ were excluded).}
\end{sidewaystable*}

\begin{sidewaystable*}
\small
\caption[Properties of the CALIFA subsample]{\label{tab:sample_califa} Properties of the 31 galaxies of the CALIFA subsample.}
   \renewcommand{\arraystretch}{0.5}
   \renewcommand{\tabcolsep}{0.08cm}
\begin{tabular}{ccccccccccc}
\hline\hline
Galaxy & Morph. Type & $z$ & $M_r$ & $R_{\rm bar}$ & $S_{\rm bar}$ & \omegabar\ & \rcor\ & \rr\ & $B/T$ & Final Sample\\
 & & & [mag] & [kpc] &  & [km s$^{-1}$ kpc$^{-1}$] & [kpc] &  &  \\
 (1) & (2) & (3) & (4) & (5) & (6) & (7) & (8) & (9) & (10) & (11)\\ 
\hline
IC 1528 & SABbc & 0.013 & $-20.57$ & 2.15$^{+0.66}_{-0.71}$ & 0.235$^{+0.002}_{-0.016}$ & $86.8\pm20.4$ & $1.63\pm0.51$ & 0.76$^{+0.14}_{-0.22}$ & 0.02 & yes\\
IC 1683 & SABb & 0.016 & $-20.73$ & 8.79$^{+0.62}_{-0.65}$ & 0.73$^{+0.07}_{-0.08}$ & $30.3\pm5.1$ & $6.33\pm2.72$ & $0.72\pm0.21$ & 0.14 & yes\\
IC 5309 & SABc & 0.014 & $-19.99$ & $1.98^{+0.89}_{-0.50}$ & $0.205^{+0.006}_{-0.019}$ & $90.6\pm26.0$ &  $1.25\pm1.01$ & 0.63$^{+0.35}_{-0.45}$ & 0.14 & yes\\
MCG-02-02-030 & SABb & 0.012 & $-20.57$ & 3.64$^{+2.27}_{-1.19}$ & 0.28$^{+0.03}_{-0.03}$ & $43.4\pm6.5$ & $4.83\pm2.16$ & 1.33$^{+0.36}_{-0.53}$ & 0.08 & yes\\
NGC 36 & SBb & 0.020 & $-21.86$ & 8.01$^{+2.02}_{-1.79}$ & 0.545 & $43.9\pm13.1$ & 5.00$^{+2.14}_{-1.55}$ & 0.6$^{+0.3}_{-0.2}$ & 0.24 & yes\\
NGC 192 & SABab & 0.014 & $-21.30$ & 11.01$^{+1.81}_{-1.45}$ & $0.83\pm0.09$ & $20.9\pm2.1$ & $11.89\pm1.85$ & 1.08$^{+0.10}_{-0.13}$ & 0.15 & yes\\
NGC 364 & EAB7 & 0.017 & $-21.28$ & 3.17$^{+0.62}_{-0.64}$ & 0.46$^{+0.01}_{-0.02}$ & $120.4\pm31.3$ & $2.63\pm1.13$ & 0.83$^{+0.22}_{-0.26}$ & 0.20 & yes\\
NGC 551 & SABbc & 0.017 &  $-20.98$ & $3.86^{+2.01}_{-2.11}$ & $0.23\pm0.07$ & $44.7\pm11.1$ & $4.52\pm2.39$ & 1.17$^{+0.39}_{-0.71}$ & 0.03 & yes\\
NGC 1645 & SB0/a & 0.016 & $-21.53$ & 5.44$^{+0.82}_{-0.44}$ & 0.692 & $65.9\pm27.6$ & 4.11$^{+2.28}_{-1.29}$ & 0.8$^{+0.4}_{-0.2}$ & 0.15 & yes\\
NGC 2449 & SABab & 0.016 & $-21.45$ & 4.59$^{+0.75}_{-0.80}$ & 0.60$^{+0.04}_{-0.03}$& $40.6\pm5.5$ & $5.84\pm0.99$ & 1.27$^{+0.11}_{-0.14}$ & 0.22 & yes\\
NGC 2553 & SABab & 0.016 & $-21.23$ & 7.68$^{+2.07}_{-1.81}$ & $0.57\pm0.01$ & $68.1\pm9.8$ & $3.95\pm0.91$ & 0.51$^{+0.08}_{-0.11}$ & 0.24 & no\\
NGC 2880 & EAB7 & 0.005 & $-20.34$ & 1.49$^{+0.71}_{-0.42}$ & 0.452$^{+0.004}_{-0.010}$ & $190.5\pm28.4$ & $1.10\pm0.36$ & 0.74$^{+0.15}_{0.19}$ & 0.51 & yes\\
NGC 3300 & SB0/a & 0.010 & $-21.17$ & 3.85$^{+0.65}_{-0.36}$ & 0.542 & $37.6\pm10.0$ & 5.86$^{+1.96}_{-1.46}$ & 1.6$^{+0.5}_{-0.4}$ & 0.10 & yes\\
NGC 3994 & SABbc & 0.010 & $-20.75$ & 1.77$^{+0.56}_{-0.47}$ & 0.382$^{+0.005}_{-0.090}$& $119.4\pm27.2$ & $1.90\pm0.67$ & 1.07$^{+0.22}_{-0.31}$ & 0.12 & yes \\
NGC 5205 & SBbc & 0.006 & $-19.65$ & 2.31$^{+0.37}_{-0.27}$ & 0.417 & $115.7\pm21.4$ & 1.48$^{+0.39}_{-0.33}$ & 0.7$^{+0.2}_{-0.1}$ & 0.06 & yes\\
NGC 5378 & SBb & 0.010 & $-20.84$ & 6.27$^{+0.97}_{-1.52}$ & 0.584 & $43.3\pm19.9$ & 4.07$^{+2.71}_{-1.65}$ & 0.6$^{+0.4}_{-0.2}$ & 0.21 & yes\\
NGC 5406 & SBb & 0.018 & $-22.25$ & 7.95$^{+0.41}_{-0.79}$ & 0.532 & $60.5\pm21.2$ & 4.15$^{+1.81}_{-1.13}$ & 0.5$^{+0.2}_{-0.1}$ & 0.12 & no\\
NGC 5947 & SBbc & 0.020 & $-21.28$ & 4.56$^{+0.54}_{-0.67}$ & 0.505 & $75.8\pm10.0$ & 2.42$^{+1.00}_{-0.96}$ & $0.5\pm0.2$ & 0.13 & no\\
NGC 5971 & SABb & 0.011 & $-20.57$ & 7.26$^{+6.12}_{-3.32}$ & 0.504$^{+0.010}_{-0.004}$ & $55.6\pm15.1$ & $4.07\pm1.96$ & 0.56$^{+0.15}_{-0.32}$ & 0.67 & no\\
NGC 6278 & SAB0/a & 0.009 & $-20.86$ & 2.84$^{+1.09}_{-0.17}$ & $0.36\pm0.04$ & $91.6\pm28.0$ & $3.05\pm1.06$ & 1.07$^{+0.26}_{-0.25}$ & 0.34 & yes\\
NGC 6427 & SAB0 & 0.011 & $-20.74$ & 1.93$^{+1.75}_{-1.05}$ & 0.63$^{+0.02}_{-0.01}$ & $46.2\pm10.4$ & $5.31\pm3.64$ & 2.76$^{+1.00}_{-1.83}$ & 0.36 & yes\\
NGC 6497 & SBab & 0.010 & $-21.72$ & 6.26$^{+0.89}_{-0.55}$ & 0.615 & $100.2\pm17.4$ & 2.34$^{+0.89}_{-0.68}$ & $0.3\pm0.1$ & 0.26 & no\\
NGC 6941 & SBb & 0.021 & $-21.57$ & 6.61$^{+0.54}_{-0.87}$ & 0.379 & $44.3\pm22.9$ & 4.45$^{+3.08}_{-1.54}$ & 0.6$^{+0.5}_{-0.2}$ & 0.09 & no\\
NGC 6945 & SB0 & 0.013 & $-21.12$ & 4.05$^{+0.66}_{-0.66}$ & 0.376 & $63.1\pm8.5$ & 3.19$^{+0.71}_{-0.61}$ & 0.8$^{+0.2}_{-0.1}$ & 0.25  & yes\\
NGC 7321 & SBbc & 0.024 & $-22.06$ & 5.75$^{+0.81}_{-0.95}$ & 0.349 & $45.5\pm13.7$ & $5.61^{+2.57}_{-1.90}$ & 1.0$^{+0.4}_{-0.3}$ & 0.05  & yes\\
NGC 7563 & SBa & 0.014 & $-21.30$ & 6.79$^{+0.75}_{-1.44}$ & 0.818 & $16.9\pm10.1$ & 12.09$^{+5.22}_{-4.21}$ & 1.0$^{+1.7}_{-0.7}$ & 0.53 & no\\
NGC 7591 & SBbc & 0.016 & $-21.50$ & 4.33$^{+0.80}_{-0.48}$ & 0.655 & $43.3\pm15.6$ & 4.23$^{+2.12}_{-1.38}$ & 1.0$^{+0.5}_{-0.3}$ & 0.19 & yes\\
UGC 3253 & SBb & 0.014 & $-20.65$ & 4.52$^{+0.37}_{-0.63}$ & 0.506 & $54.2\pm10.8$ & 3.40$^{+0.91}_{-0.77}$ & 0.7$^{+0.2}_{-0.2}$ & 0.07 & yes\\
UGC 3944 & SABbc & 0.013 & $-20.03$ & 1.87$^{+0.88}_{-0.63}$ & 0.27$^{+0.05}_{-0.02}$& $61.8\pm21.6$ & $2.39\pm11.62$ & 1.28$^{+3.80}_{-5.66}$ & 0.00 & yes\\
UGC 8231 & SABd & 0.008 & $-18.71$ & 2.30$^{+0.47}_{-0.70}$ & 0.24$^{+0.04}_{-0.04}$& $58.3\pm30.8$ & $2.33\pm5.29$ & 1.01$^{+1.61}_{-1.99}$ & 0.00 & no\\
UGC 12185 & SBb & 0.022 & $-21.30$ & 8.97$^{+3.67}_{-1.85}$ & 0.710 & $22.6\pm4.5$ & 9.54$^{+5.43}_{-4.02}$ & 1.2$^{+0.6}_{-0.5}$ & 0.20 & yes\\
\hline
\end{tabular}
\tablefoot{(1) Galaxy name. (2) Morphological classification from CALIFA \citep{Walcher2014}. (3) Redshift from SDSS-DR14 \citep{Abolfathi2018}. (4) Absolute SDSS $r$-band magnitude obtained as described in Sect.~\ref{sec:sample}. (5) Bar radius obtained as described in Sect.~\ref{sec:abar}. (6) Bar strength obtained as described in Sect.~\ref{sec:sbar}. (7) Bar pattern speed obtained as described in Sect.~\ref{sec:pattern}. (8) Bar corotation radius obtained as described in Sect.~\ref{sec:rcr_r}. (9) Bar rotation rate obtained as described in Sect.~\ref{sec:rcr_r}. (10) Bulge-to-total luminosity ratio provided by \cite{MendezAbreu2017}. (11) Inclusion in the final analysed sample (galaxies with no were excluded).}
\end{sidewaystable*}

\clearpage
\onecolumn
\begin{landscape}
\centering
\normalsize
\begin{longtable}{cccccccccccc}
\caption{\label{tab:sample_manga} Properties of the 55 galaxies of the MaNGA subsample.}\\
\hline\hline
Galaxy & Morph. Type & $z$ & $M_r$ & $R_{\rm bar}$ & $S_{\rm bar}$ & \omegabar\ & \rcor\ & \rr & $B/T$ (Ref.) & Ref. & Final sample\\
 & & & [mag] & [kpc] &  & [km s$^{-1}$ kpc$^{-1}$] & [kpc] & & & & \\
 (1) & (2) & (3) & (4) & (5) & (6) & (7) & (8) & (9) & (10) & (11) & (12)\\ 
\hline
7495-12704 & SBbc & 0.029 & $-21.40$ & 4.70$^{+0.69}_{-0.63}$ & 0.37 & 30.3$^{+3.6}_{-2.8}$ & 6.70$^{+1.13}_{-1.00}$ & 1.43$^{+0.33}_{-0.28}$ & ... & 1 & yes \\
7962-12703 & SBab & 0.048 & $-22.33$ & 16.11$^{+3.70}_{-3.00}$ & 0.65 & 27.8$^{+0.9}_{-0.7}$ & 9.40$^{+1.20}_{-1.10}$ & 0.58$^{+0.16}_{-0.12}$ & ... & 1 & no \\	
7990-3704 & SB0 & 0.029 & $-20.15$ & 2.37$^{+0.30}_{-0.42}$ & 0.29 & 79.7$^{+25.4}_{-25.2}$ & 1.88$^{+0.91}_{-0.49}$ & 0.84$^{+0.42}_{-0.26}$ & ... & 1 & yes \\
7990-9101 & SBc & 0.028 & $-19.77$ & 4.03$^{+0.64}_{-1.11}$ & 0.37 & 15.5$^{+5.0}_{-5.9}$ & 7.72$^{+4.91}_{-2.05}$ & 2.15$^{+1.39}_{-0.77}$ & ... & 1 & yes\\
7990-12704 & SBa & 0.026 & $-21.12$ & $7.01\pm0.47$ & 0.62$^{+0.06}_{-0.05}$ & $33.3^{+3.8}_{-7.2}$ & 5.27$^{+2.81}_{-2.32}$ & 0.76$^{+0.41}_{-0.33}$ & ... & 2 & yes\\
7992-6104 & SBc & 0.027 & $-20.31$ & 5.11$^{+0.91}_{-0.79}$ & 0.80 & 27.1$^{+1.9}_{-1.7}$ & 4.65$^{+0.68}_{-0.62}$ & 0.91$^{+0.22}_{-0.18}$ & ... & 1 & yes\\
8082-6102 & SB0 & 0.024 & $-21.46$ & 3.81$^{+0.50}_{-0.50}$ & 0.59 & 50.8$^{+23.0}_{-19.4}$ & 4.66$^{+3.81}_{-1.40}$ & 1.28$^{+0.97}_{-0.44}$ & ... & 1 & yes\\
8083-6102 & SBa & 0.036 & $-21.62$ & 5.28$^{+1.13}_{-1.21}$ & 0.63 & 12.4$^{+4.8}_{-3.2}$ & 23.25$^{+14.50}_{-5.74}$ & 4.73$^{+2.88}_{-1.61}$ & ... & 1 & yes\\
8083-12704 & SBbc & 0.023  & $-21.03$ & 3.09$^{+0.47}_{-0.52}$ & 0.27 & 85.0$^{+50.0}_{-82.1}$ & 1.12$^{+1.59}_{-0.52}$ & 0.39$^{+0.51}_{-0.19}$ & 0.03 & 1 & no\\
8133-3701 & SBb & 0.044 & $-20.10$ & 3.88$^{+0.83}_{-1.02}$ & 0.48 & 41.8$^{+6.3}_{-8.9}$ & 3.32$^{+0.74}_{-0.65}$ & 0.88$^{+0.35}_{-0.24}$ & 0.08 & 1 & yes\\
8134-6102 & SB0a & 0.032 & $-21.40$ &  7.95$^{+1.90}_{-1.36}$ & 0.74 & 23.0$^{+4.7}_{-3.8}$ & 12.37$^{+3.53}_{-2.31}$ & 1.56$^{+0.56}_{-0.41}$ & 0.19 & 1 & yes\\
8137-9102 & SBb  & 0.031 & $-21.07$ &  7.65$^{+0.67}_{-1.26}$ & 0.62 & 33.1$^{+4.4}_{-8.8}$ & 3.86$^{+0.93}_{-0.80}$ & 0.53$^{+0.15}_{-0.13}$ & ... & 1 & no\\
8140-12701 & SBa  & 0.029 & $-20.61$ &  5.86$^{+0.98}_{-0.73}$ & 0.68 & 39.5$^{+8.3}_{-6.1}$ & 4.34$^{+1.04}_{-0.92}$ & 0.73$^{+0.22}_{-0.17}$ & ... & 1 & yes\\
8140-12703 & SBb  & 0.032 & $-21.87$ &  7.31$^{+1.37}_{-1.43}$ & 0.37 & 28.2$^{+11.4}_{-7.9}$ & 7.31$^{+4.78}_{-1.78}$ & 1.07$^{+0.68}_{-0.35}$ & ... & 1 & yes\\
8243-6103 & SB0 & 0.032 & $-21.65$ &  4.77$^{+0.40}_{-0.67}$ & 0.70 & 21.3$^{+16.8}_{-15.4}$ & 14.12$^{+27.90}_{-6.05}$ & 3.31$^{+5.80}_{-1.58}$ & 0.16 & 1 & no\\
8243-12704 & SBbc & 0.024 & $-20.56$ & $3.27\pm0.63$ & $0.22^{+0.12}_{-0.11}$ & $34.2^{+17.8}_{-11.1}$ & $4.54^{+3.40}_{-2.32}$ & $1.39^{+1.07}_{-0.72}$ & ... & 2 & no\\
8244-3703 & SB0  & 0.048 & $-21.03$ &  4.30$^{+0.41}_{-0.72}$ & 0.38 & 73.0$^{14.2}_{-12.9}$ & 2.77$^{+0.72}_{-0.51}$ & 0.67$^{+0.20}_{-0.15}$ & ... & 1 & yes\\
8247-3701 & SB0a & 0.025 & $-20.59$ & 2.53$^{+0.38}_{-0.70}$ & 0.40 & 22.1$^{+5.4}_{-11.2}$ & 5.38$^{+2.10}_{-1.72}$ & 2.27$^{+1.15}_{-0.81}$ & ... & 1 & yes\\
8249-6101 & SBc  & 0.027 & $-20.27$ &  7.42$^{+0.69}_{-0.92}$ & 1.13 & 30.5$^{+2.8}_{-3.2}$ & 4.31$^{+0.69}_{-0.63}$ & 0.59$^{+0.12}_{-0.10}$  & ... & 1 & no\\
8254-9101 & SBa  & 0.025 & $-21.78$ &  6.91$^{+0.55}_{-0.60}$ & 0.51 & 48.5$^{+26.0}_{-44.1}$ & 6.41$^{+12.77}_{-2.30}$ & 0.96$^{+1.82}_{-0.36}$ & ... & 1 & no\\
8256-6101 & SBa  & 0.025 & $-20.79$ &  5.02$^{+0.75}_{-0.48}$ & 0.64 & $36.2^{+27.8}_{-31.9}$ & 5.29$^{+11.05}_{-2.24}$ & 1.10$^{+2.12}_{-0.51}$ & 0.45 & 1 & no\\
8257-3703 & SBb & 0.025 & $-20.34$ &  3.98$^{+1.13}_{-0.97}$ & 0.76 & $50.1\pm2.4$ & $3.87\pm0.48$ & 0.97$^{+0.34}_{-0.24}$ & 0.07 & 1 & yes\\
8257-6101 & SBc & 0.029 & $-20.86$ &  2.57$^{+0.25}_{-0.31}$ & 0.20 & $48.3^{+23.9}_{-25.9}$ & 3.45$^{+3.45}_{-1.25}$ & 1.42$^{+1.28}_{-0.56}$ & ... & 1 & yes\\
8312-12702 & SBc & 0.032 & $-21.24$ &  6.58$^{+1.22}_{-1.42}$ & 0.63 & $34.9^{+4.8}_{-5.6}$ & 4.07$^{+0.88}_{-0.68}$ & 0.63$^{+0.22}_{-0.14}$ & ... & 1 & yes\\
8312-12704 & SBb  & 0.030 & $-21.00$ &  7.00$^{+1.38}_{-1.75}$ & 0.60 & $14.4^{+5.1}_{-4.4}$ & 8.69$^{+4.94}_{-2.25}$ & 1.33$^{+0.80}_{-0.44}$ & 0.06 & 1 & yes\\
8313-9101 & SBb & 0.039 & $-21.87$ &  4.39$^{+0.73}_{-1.46}$ & 0.24 & $0.8^{+11.4}_{-23.5}$ & 21.62$^{+59.65}_{-12.76}$ & 6.31$^{+14.20}_{-4.12}$ & ... & 1 & no\\
8317-12704 & SBa  & 0.054 & $-22.68$ & 11.88$^{+1.15}_{-1.73}$ & 0.71 & $12.2^{+2.9}_{-2.8}$ & 27.69$^{+8.88}_{-6.00}$ & 2.43$^{+0.83}_{-0.61}$ & 0.15 & 1 & yes\\
8318-12703  & SBb  & 0.039 & $-22.21$ &  6.53$^{+1.76}_{-2.01}$ & 0.44 & $28.7^{+5.8}_{-7.9}$ & 8.37$^{+3.10}_{-1.76}$ & 1.35$^{+0.74}_{-0.43}$ & ... & 1 & yes\\
8320-6101   & SBb  & 0.027 & $-20.37$ &  3.80$^{+0.81}_{-0.46}$ & 0.43 & $27.2^{+5.4}_{-4.8}$ & 6.90$^{+1.67}_{-1.38}$ & 1.78$^{+0.54}_{-0.44}$ & ... & 1 & yes\\
8326-3704   & SBa  & 0.026 & $-20.25$ &  4.07$^{+0.51}_{-0.85}$ & 0.45 & $15.0^{+17.0}_{-39.0}$ & 6.62$^{+16.29}_{-4.47}$ & 1.90$^{+4.12}_{-1.35}$ & ... & 1 & no\\
8326-6102   & SBb  & 0.070 & $-22.06$ &  8.00$^{+0.89}_{-1.48}$ & 0.56 & $19.0^{+8.3}_{-13.3}$ & 12.15$^{+10.37}_{-14.89}$ & 1.62$^{+1.35}_{-0.71}$ & ... & 1 & yes\\
8330-12703  & SBbc & 0.027 & $-20.67$ &  5.80$^{+0.70}_{-0.81}$ & 0.31 & $44.9^{+4.1}_{-3.7}$ & 3.07$^{+0.52}_{-0.41}$ & 0.54$^{+0.12}_{-0.10}$ & ... & 1 & no\\
8335-12701  & SBb  & 0.063 & $-21.66$ & 12.05$^{+4.82}_{-4.42}$ & 0.60 & $7.9^{+4.5}_{-2.7}$ & 29.47$^{+16.47}_{-10.45}$ & 2.53$^{+2.13}_{-1.13}$ & ... & 1 & no\\
8341-12704 & SBbc & 0.031 & $-21.49$ & $5.11\pm 0.79$ & 0.50$^{+0.07}_{-0.08}$ & $25.7^{+6.4}_{-7.3}$ & $4.71^{+2.94}_{-1.81}$ & $0.92^{+0.58}_{-0.36}$ & ... & 2 & yes\\
8439-6102 & SBab & 0.034 & $-21.64$ & 5.36$^{+1.45}_{-1.52}$ & 0.53 & $53.6\pm1.5$ & 3.84$^{+0.43}_{-0.51}$ & 0.71$^{+0.29}_{-0.17}$ & 0.16 & 1 & yes\\
8439-12702 & SBa  & 0.027 & $-21.57$ & $6.23\pm0.63$ & 0.46 & $30.8^{+4.2}_{-5.1}$ & 7.73$^{+1.79}_{-1.27}$ & 1.25$^{+0.31}_{-0.24}$ & ... & 1 & yes\\
8440-12704 & SBb  & 0.027 & $-21.12$ &  3.26$^{+0.70}_{-0.64}$ & 0.43 & $35.9^{+7.5}_{-4.3}$ & 5.81$^{+1.22}_{-1.10}$ & 1.79$^{+0.59}_{-0.45}$ & ... & 1 & yes\\
8447-6101 & SBb  & 0.075 & $-22.89$ & 14.65$^{+1.11}_{-1.75}$ & 0.30 & $37.6^{+7.5}_{-11.3}$ & 9.39$^{+2.87}_{-2.23}$ & $0.66^{+0.21}_{-0.17}$ & ... & 1 & yes\\
8452-3704 & SBc  & 0.025 & $-19.97$ & 2.16$^{+0.92}_{-0.65}$ & 0.21 & $76.2^{+48.0}_{-51.2}$ & 2.06$^{+3.03}_{-0.87}$ & 1.07$^{+1.39}_{-0.58}$ & ... & 1 & no\\
8452-12703 & SBb & 0.061 & $-22.83$ & 9.19$^{+1.94}_{-2.98}$ & 0.38 & $42.2^{+6.0}_{-5.6}$ & 5.05$^{+1.04}_{-0.91}$ & 0.57$^{+0.28}_{-0.15}$ & ... & 1 & yes\\
8453-12701 & SABc & 0.026 & $-20.58$ & $3.69\pm0.21$ & $0.58\pm0.01$ & $26.9^{+14.3}_{-2.8}$ & $4.03^{+2.03}_{-1.66}$ & $1.08^{+0.55}_{-0.45}$ & ... & 2 & no \\
8481-12701 & SBa & 0.067 & $-21.91$ & 8.15$^{+1.12}_{-1.41}$ & 0.65 & $40.2^{+10.2}_{-7.1}$ & 6.60$^{+2.25}_{-1.26}$ & 0.85$^{+0.31}_{-0.21}$ & ... & 1 & yes\\
8482-9102 & SBb & 0.058 & $-21.59$ & 6.84$^{+0.73}_{-1.22}$ & 0.41 & $15.3^{+5.9}_{-3.8}$ & 14.65$^{+8.67}_{-3.54}$ & 2.34$^{+1.30}_{-0.70}$ & 0.13 & 1 & yes\\
8482-12703 & SBbc & 0.050 & $-22.21$ & 6.15$^{+1.15}_{-1.36}$ & 0.41 & $41.7^{+15.6}_{-15.9}$ & 3.96$^{+2.29}_{-1.15}$ & 0.68$^{+0.42}_{-0.24}$ & ... & 1 & yes\\
8482-12705 & SBb & 0.042 & $-22.06$ & 8.24$^{+0.88}_{-1.32}$ & 0.32 & $12.9^{+6.1}_{-8.2}$ & 18.42$^{+25.52}_{-6.31}$ & 2.47$^{+2.96}_{-0.98}$ & ... & 1 & yes\\
8486-6101 & SBc & 0.059 & $-21.57$ & 5.83$^{+1.61}_{-2.11}$ & 0.59 & $18.8^{+3.9}_{-4.7}$ & 10.05$^{+3.47}_{-2.11}$ & 1.84$^{+1.10}_{-0.61}$ & 0.00 & 1 & yes\\
8548-6102 & SBc & 0.048 & $-20.83$ & 7.05$^{+1.61}_{-1.51}$ & 0.98 & $35.2^{+5.5}_{-3.9}$ & 4.53$^{+0.91}_{-0.70}$ & $0.65^{+0.21}_{-0.16}$ & 0.00 & 1 & yes\\
8548-6104 & SBc & 0.048 & $-20.47$ & 4.94$^{+0.71}_{-0.91}$ & 0.49 & $23.4^{+4.3}_{-4.5}$ & 7.56$^{+1.92}_{-1.51}$ & 1.56$^{+0.51}_{-0.37}$ & ... & 1 & yes\\
8549-12702 & SBb & 0.043 & $-22.03$ & 5.42$^{+0.83}_{-0.46}$ & 0.49 & $76.0^{+17.0}_{-23.5}$ & 3.21$^{+1.01}_{-0.83}$ & 0.58$^{+0.20}_{-0.16}$ & ... & 1 & no\\
8588-3701 & SBb & 0.130 & $-22.88$ & 13.67$^{+1.91}_{-2.19}$ & 0.46 & $45.3^{+12.7}_{-12.6}$ & 5.74$^{+2.46}_{-1.37}$ & 0.44$^{+0.19}_{-0.13}$ & ... & 1 & no\\
8601-12705 & SBc & 0.030 & $-21.21$ & 4.07$^{+1.00}_{-0.81}$ & 0.40 & $23.5^{+4.8}_{-2.1}$ & $7.33\pm1.32$ & 1.78$^{+0.60}_{-0.44}$ & ... & 1 & yes\\
8603-12703 & SBa & 0.030 & $-21.04$ & 7.65$^{+0.57}_{-0.82}$ & 0.30 & $25.2^{+9.3}_{-11.7}$ & 5.82$^{+3.54}_{-1.90}$ & 0.79$^{+0.46}_{-0.28}$ & ... & 1 & yes\\
8604-12703 & SBab & 0.031 & $-21.67$ & 6.59$^{+0.83}_{-1.41}$ & 0.50 & $16.4^{+7.9}_{-20.1}$ & 13.57$^{+26.11}_{-4.54}$ & 2.45$^{+3.94}_{-1.06}$ & 0.14 & 1 & yes\\
8612-6104 & SBb & 0.036 & $-21.83$ & 6.12$^{+1.64}_{-1.12}$ & 0.56 & $104.3^{+11.9}_{-12.9}$ & $1.79\pm0.30$ & 0.29$^{+0.09}_{-0.07}$ & ... & 1 & no\\
8612-12702 & SBc & 0.063 & $-22.60$ & 7.15$^{+1.19}_{-1.06}$ & 0.30 & $41.2^{+33.2}_{-23.6}$ & 5.03$^{+6.35}_{-2.25}$ & 0.74$^{+0.87}_{-0.36}$ & ... & 1 & no\\
\hline
\end{longtable}
\tablefoot{(1) Galaxy name. (2) Morphological classification from \citet{GarmaOehmichen2019} and \citet{Guo2019}. (3) Redshift from MaNGA \citep{bundy2015}. (4) Absolute SDSS $r$-band magnitude obtained as described in Sect.~\ref{sec:sample}. (5) Bar radius obtained as described in Sect.~\ref{sec:abar}. (6) Bar strength obtained as described in Sect.~\ref{sec:sbar}. (7) Bar pattern speed obtained as described in Sect.~\ref{sec:pattern}. (8) Bar corotation radius obtained as described in Sect.~\ref{sec:rcr_r}. (9) Bar rotation rate obtained as described in Sect.~\ref{sec:rcr_r}. (10) Bulge-to-total luminosity ratio provided by \cite{Kruk2018}. (11) Reference paper for the direct measurement of the bar pattern speed: 1 = \cite{Guo2019}, 2 = \cite{GarmaOehmichen2019}. (12) Inclusion in the final analysed sample (galaxies with no were excluded).}
\end{landscape}
\clearpage
\twocolumn

\section{Relations among the bar parameters in ETBGs and LTBGs}
\label{appendix:b}

We plot the relations after splitting the final sample between ETBGs and LTBGs.

\begin{figure*}[!t]
    \centering
    \includegraphics[scale=0.7]{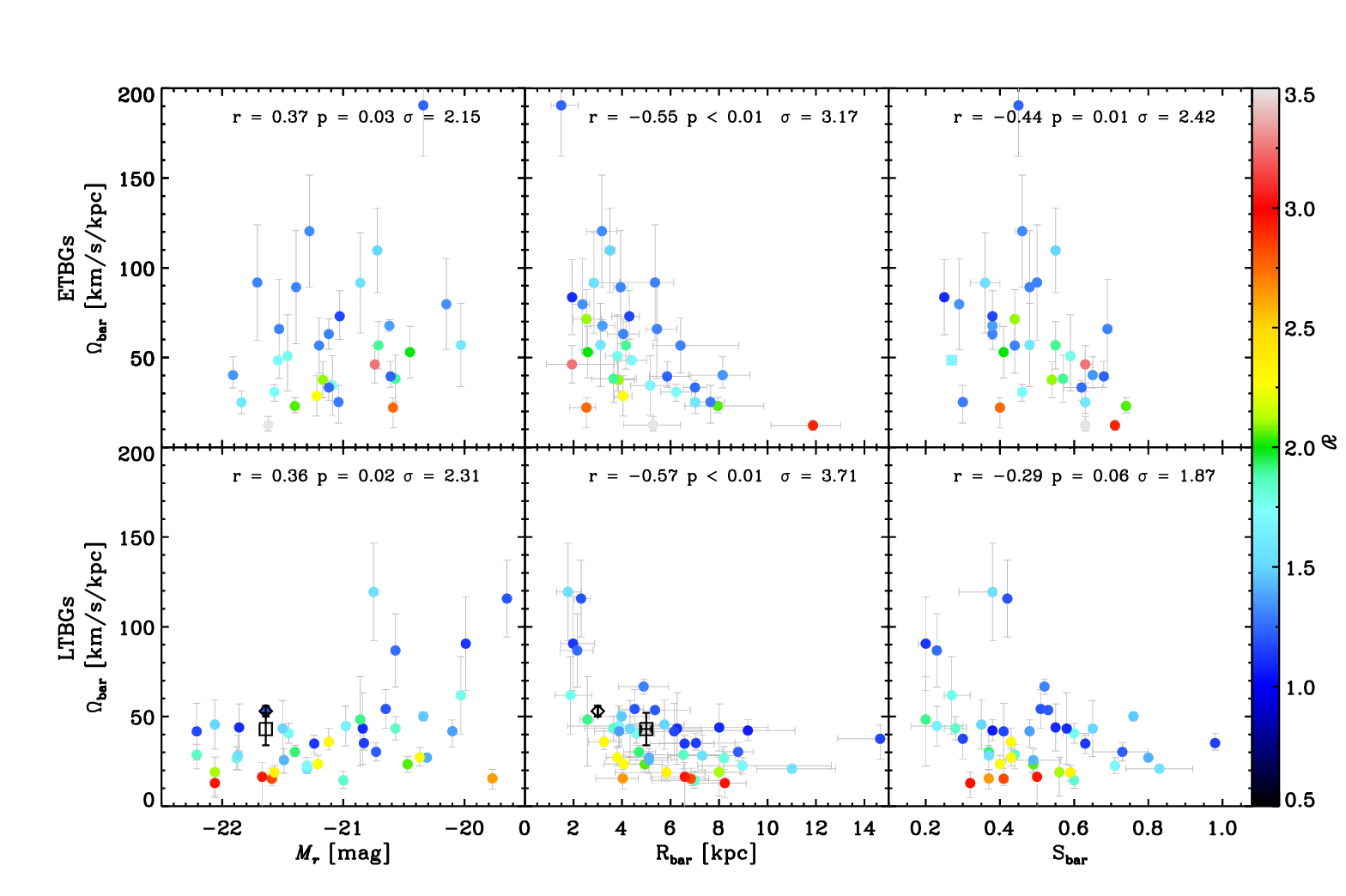}
    \caption{{\em Top panels:} relations between the bar pattern speed \omegabar\ and the SDSS $r$-band absolute magnitude $M_r$, bar radius \rbar, and bar strength \sbar\ for the subsample of 34 ETBGs. {\em Bottom panels:} same as above but for the subsample of 43 LTBGs. The Spearman rank correlation $r$, the two-sided significance $p$, and number of $\sigma$ from the null-hypothesis are given in each panel. The points are colour-coded according to the value of \rr. Results obtained using different bands with respect to the SDSS $r$-band are represented with a square symbol. The results for the Milky Way are shown both for the short bar case ({\em black open diamond}) and long bar case ({\em black open square}).}
    \label{fig:omega_relation_e_l}
\end{figure*}

\begin{figure*}[!t]
    \centering
\includegraphics[scale=0.7]{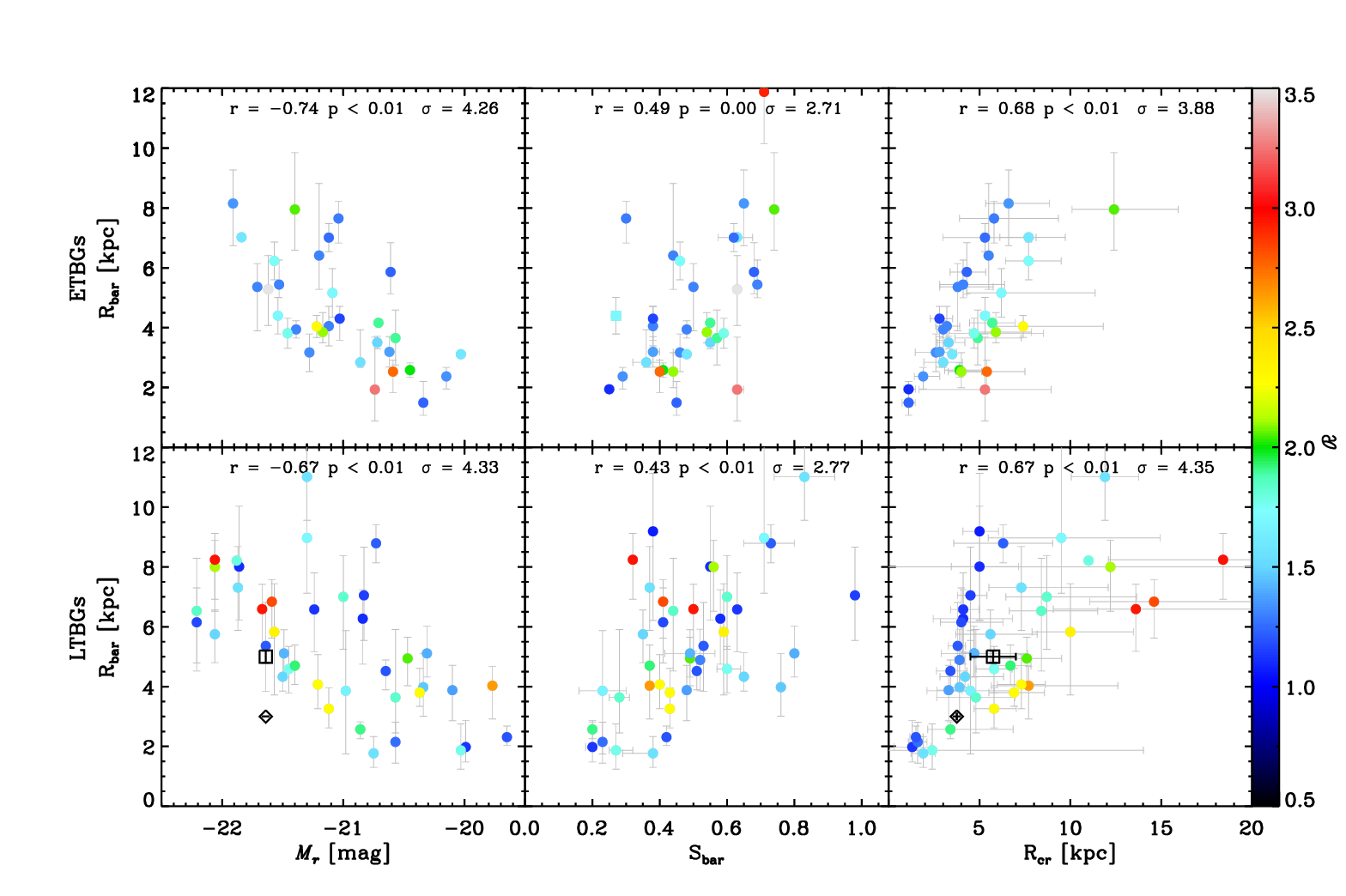}
    \caption{{\em Top panels:} relations between the bar radius \rbar\ and the absolute SDSS $r$-band magnitude $M_r$, bar strength \sbar, and corotation radius \rcor\ for the subsample of 34 ETBGs. {\em Bottom panels:} same as above but for the subsample of 43 LTBGs. The Spearman rank correlation $r$, two-sided significance $p$, and number of $\sigma$ from the null-hypothesis are given in each panel. The points are colour-coded according to the value of \rr. Results obtained using different bands with respect to the SDSS $r$-band are represented with a square symbol. The results for the Milky Way are shown both for the short bar case ({\em black open diamond}) and long bar case ({\em black open square}).}
    \label{fig:rbar_relation_e_l}
\end{figure*}

\section{Scaled relations involving the sizes of galaxies}
\label{appendix:c}

We plot the relations shown in the third panel of Fig.~\ref{fig:omega_relation} and in Fig.~\ref{fig:rbar_relation} after scaling the sizes of galaxies with \rpetro.

\begin{figure*}[!t]
    \centering
    \includegraphics[scale=0.7]{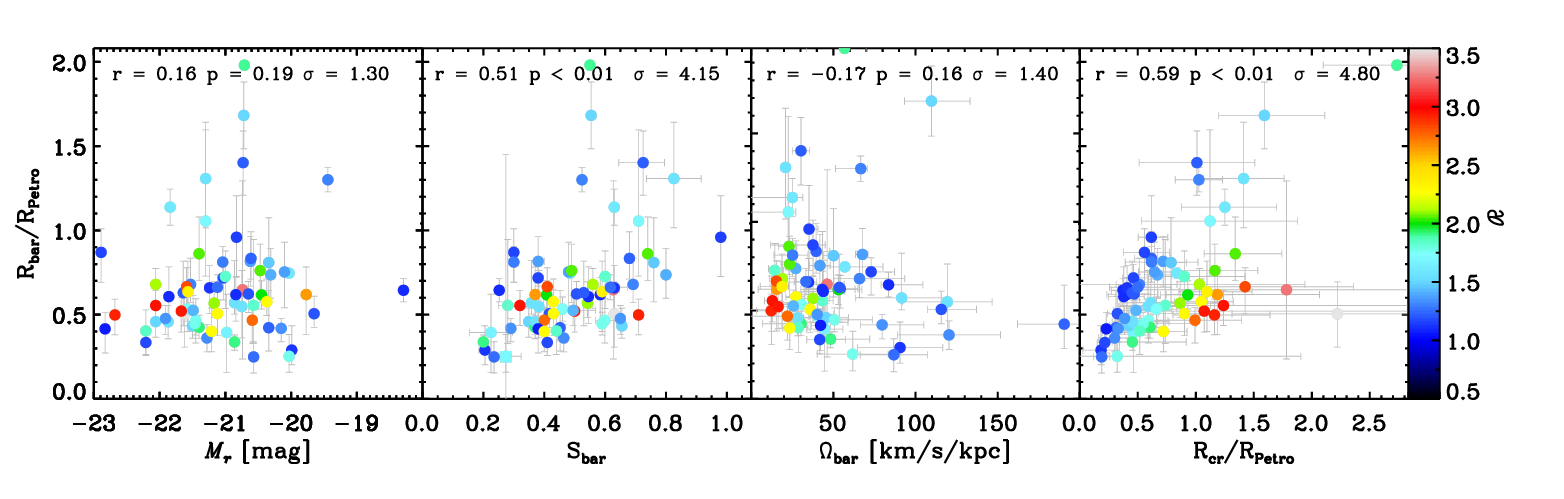}
    \caption{Relations between the bar radius \rbar\ normalised by \rpetro, and the absolute SDSS $r$-band magnitude $M_r$, bar strength \sbar, bar pattern speed \omegabar, and corotation radius \rcor\ normalised by \rpetro, for the subsample of 66 galaxies for which we have \rpetro. The Spearman rank correlation $r$, the two-sided significance $p$, and number of $\sigma$ from the null-hypothesis are given in each panel. The points are colour-coded according to the value of \rr. Results obtained using different bands with respect to the SDSS $r$-band are represented with a square symbol. The results for the Milky Way are shown both for the short bar case ({\em black open diamond}) and long bar case ({\em black open square}).}
    \label{fig:scaled_relation}
\end{figure*}


\end{appendix}

\label{lastpage}
\end{document}